\documentclass[aip,jcp,reprint,floatfix]{revtex4-1}
\pdfoutput=1

\usepackage{siunitx} 
\usepackage{amsmath} 
\usepackage{amssymb}
\usepackage{mathrsfs}
\usepackage{mathtools}
\usepackage{textcomp}
\usepackage{stmaryrd}
\usepackage{empheq}
\usepackage{esint}

\usepackage[shortlabels]{enumitem}
\usepackage{graphicx} 

\usepackage{multirow}
\usepackage{booktabs} 
\usepackage{hyperref} 

\DeclareMathAlphabet{\mathpzc}{OT1}{pzc}{m}{it}

\DeclareSIUnit{\molar}{M}
\DeclareSIUnit{\bp}{bp}

\AtBeginDocument
{
  \heavyrulewidth=.08em
  \lightrulewidth=.05em
  \cmidrulewidth=.03em
  \belowrulesep=.65ex
  \belowbottomsep=0pt
  \aboverulesep=.4ex
  \abovetopsep=0pt
  \cmidrulesep=\doublerulesep
  \cmidrulekern=.5em
  \defaultaddspace=.5em
}

\begin{document}
\title{Chiral shape fluctuations and the origin of chirality in cholesteric phases of DNA origamis}
\date{\today}
\author{Maxime M.C. Tortora}
\thanks{Current address: Laboratory of Biology and Modeling of the Cell, \'Ecole Normale Sup\'erieure de Lyon, 46, all\'ee d’Italie, 69364 Lyon Cedex 07, France}
\thanks{correspondence to \href{mailto:maxime.tortora@ens-lyon.fr}{maxime.tortora@ens-lyon.fr}}
\affiliation{Physical and Theoretical Chemistry Laboratory, Department of Chemistry, University of Oxford, South Parks Road, Oxford OX1 3QZ, United Kingdom}
\author{Garima Mishra}
\affiliation{Department of Physics, Indian Institute of Technology Kanpur, Kanpur 208016, India\looseness=-1}
\author{Domen Pre\v{s}ern}
\affiliation{Physical and Theoretical Chemistry Laboratory, Department of Chemistry, University of Oxford, South Parks Road, Oxford OX1 3QZ, United Kingdom}
\author{Jonathan P.K. Doye}
\affiliation{Physical and Theoretical Chemistry Laboratory, Department of Chemistry, University of Oxford, South Parks Road, Oxford OX1 3QZ, United Kingdom}

\keywords{Liquid crystals; colloids; self-assembly; chirality; DNA origamis.}

\begin{abstract}
Lyotropic cholesteric liquid crystal phases are ubiquitously observed in biological and synthetic polymer solutions, characterized by a complex interplay between thermal fluctuations, entropic and enthalpic forces. The elucidation of the link between microscopic features and macroscopic chiral structure, and of the relative roles of these competing contributions on phase organization, remains a topical issue. Here we provide theoretical evidence of a novel mechanism of chirality amplification in lyotropic liquid crystals, whereby phase chirality is governed by fluctuation-stabilized helical deformations in the conformations of their constituent molecules. Our results compare favorably to recent experimental studies of DNA origami assemblies and demonstrate the influence of intra-molecular mechanics on chiral supra-molecular order, with potential implications for a broad class of experimentally-relevant colloidal systems.
\end{abstract}

\maketitle

\section*{Introduction}

Linking the microscopic features of molecular building blocks to the material properties of their self-assembled macroscopic phases constitutes one of the overarching goals of modern colloidal science. The fascinating ability of colloidal systems to spontaneously form intricate, organized structures in the absence of external human intervention has inspired a considerable body of work over the last decades, fostered by rapid experimental progress in the synthesis of micro- and nano-sized particles with complex shapes and tunable interactions.\cite{Glot07} Colloidal self-assembly has thus emerged as a most promising route towards the bottom-up fabrication of functional nano-materials, with potential applications spanning the fields of catalysis, energy harvesting and drug delivery.\cite{Stein13, Beij12} However, the hierarchical nature of the colloidal assembly process --- which involves the gradual propagation of orientational and/or positional order from molecular to macroscopic level --- is generally challenging to rationalize and control, owing to both the diversity of physico-chemical forces at play and the wide difference in length-scales between elementary building blocks and super-molecular structures.\cite{Elac17}
\par
In this context, the question of the role of molecular chirality on colloidal organization has proven to be of singularly long-standing interest, from the point of view of both practical applications and fundamental research. The control of the chirality of a macroscopic material, along with the elucidation of its microscopic bases, has far-reaching implications ranging from pharmaceutical synthesis and photonics engineering\cite{Wang13, Hent17} to the understanding of the origins of biological homochirality.\cite{Bern67} In particular, the self-assembly of chiral molecular units into helical super-structures underlies the formation of the basic molecules of life --- from the double-helical ordering of nucleotides in DNA to the $\alpha$-helical arrangement of amino-acids in protein secondary structures --- and governs their remarkable ability to further organize into higher-order helical assemblies, such as helix-bundle proteins and protein-DNA complexes, which are essential to vital biological functions.\cite{Yash16}
\par
In colloidal systems, the most frequent manifestation of this so-called \textit{chirality amplification} process lies in the lyotropic cholesteric liquid crystal (LChLC) phase, observed in solutions of many common chiral (bio)polymers in both \textit{in vivo} and \textit{in vitro} environments. The macroscopic breaking of mirror symmetry in LChLCs arises from the periodic rotation of the direction of local molecular alignment about a fixed normal axis as one passes through the sample, and may be fully quantified by the spatial period of this helical arrangement --- termed the \textit{cholesteric pitch}. A remarkable feature of cholesterics, whose original discovery in 1888 is generally hailed as the birth of liquid crystal science,\cite{Rein88} is the exquisite sensitivity of their pitch to subtle changes in the assembly conditions and chemical structure of their constituent particles. This delicate dependence has been studied in considerable detail in a variety of model systems, ranging from DNA duplex\cite{Stan05} and filamentous virus suspensions\cite{Grel03} to biologically-relevant collagen assemblies,\cite{DeSa11} and forms the basis of an impressive array of potential applications in such diverse fields as cryptography, smart textiles and physico-chemical sensors.\cite{Schw18}
\par
Despite notable recent advances,\cite{deMi16} the striking complexity and heterogeneity of the reported experimental phase behaviors has so far largely eluded attempts to resolve their microscopic underpinnings. While theoretical studies of simple particle models have uncovered a few general features of cholesteric organization, such as the non-trivial link between molecular and phase chirality,\cite{Wens11,Frez14,Duss16} such investigations have been hindered by the strongly multi-scale nature of the problem, as the cholesteric pitch of most common LChLCs usually lies in the micro- to millimeter range --- several orders of magnitude larger than the typical molecular dimensions --- which renders direct atomistic simulations largely impractical. Conversely, on the experimental side, the systematic analysis of the relationship between molecular structure and phase organization has been limited by the difficulties involved in the scalable fabrication of colloidal particles with addressable chirality. The establishment of a quantitative connection between molecular chirality and macroscopic helicity in LChLCs thus generally remains a major challenge of soft condensed-matter physics, with broad consequences for their rational applications as bio-inspired multifunctional materials\cite{Lage14,Wang18} and for our fundamental understanding of the ubiquitous occurrences of LChLC order in living matter.\cite{Mito17}
\par
Substantial progress in this direction has been recently achieved by exploiting the synergy between colloidal science and DNA origami technology, through which the LChLC organization of self-assembled origami filaments demonstrated the possibility to tune the micron-scale pitch of the bulk phase via the direct control of single-particle structure at the nanometer level.\cite{Siav17} Through the conjunction of a well-established coarse-grained model of DNA with a classical molecular field theory of LChLCs, we here present a rigorous theoretical analysis of these experimental developments by assessing the detailed influence of particle mechanical properties and thermodynamic state on their ordering behavior, without the use of any adjustable parameters. These developments further enable us to demonstrate for the first time the importance of intra-molecular fluctuations on chiral supra-molecular organization, and thus provide the first tangible evidence of a long-postulated, general mechanism of chirality amplification in biopolymer solutions.\cite{Grel03}

\begin{figure}[]
  \includegraphics[width=\columnwidth]{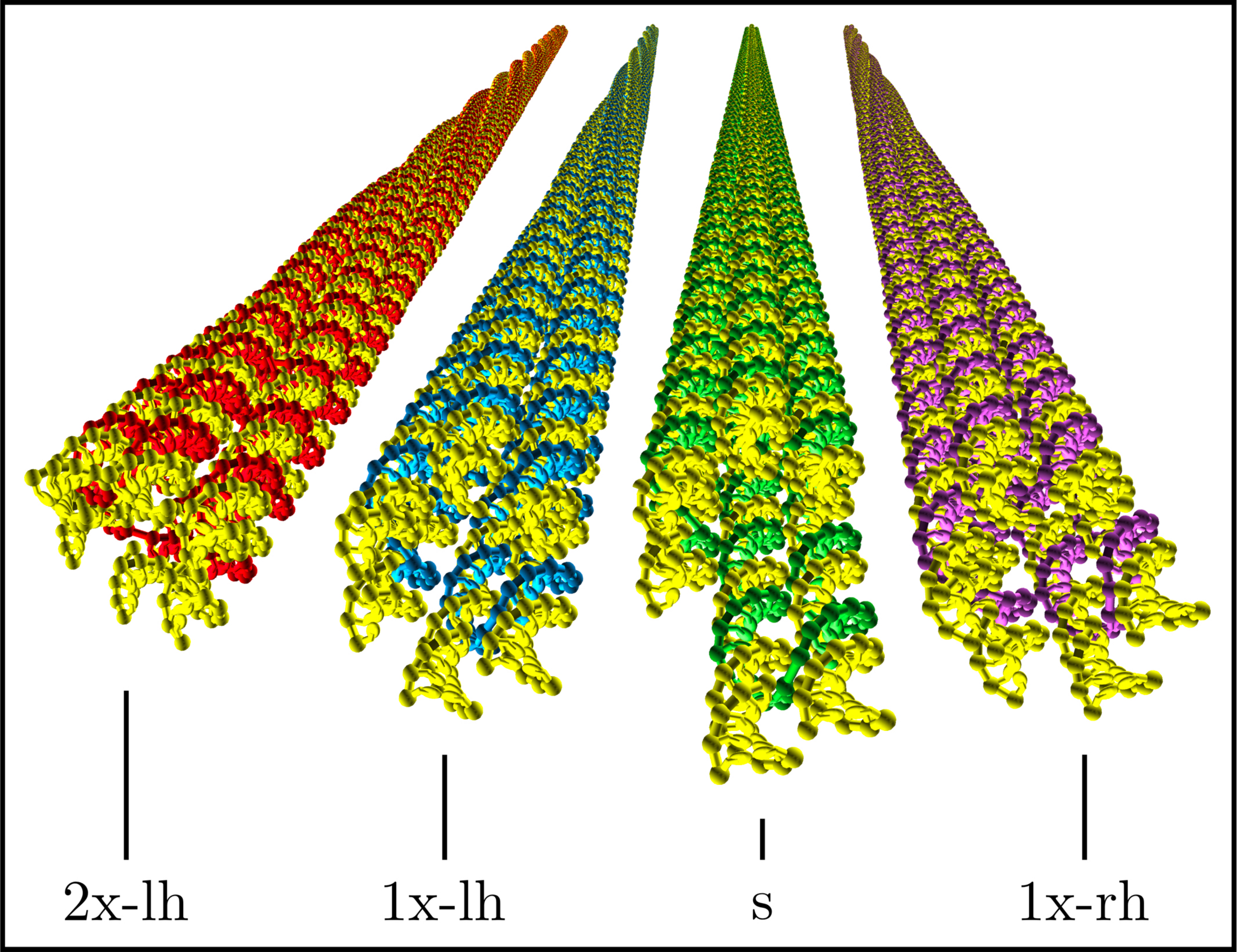}
  \caption{\label{fig1}\textbf{Ground-state origami conformations.} Each origami is comprised of a 7,560-nucleotide scaffold strand, colored in yellow, and of $\sim 200$ shorter staple strands, colored independently for each origami variant, whose designed binding locations determine the filament ground-state structure. The equilibrium axial twist of the conformations is obtained by elastic energy minimization using a continuum DNA model.\cite{Kim17} The nucleotide-level depiction corresponds to the finer-grained representation of the oxDNA model,\cite{Snod15} as employed in all mechanical calculations throughout the paper.}
\end{figure}

\section*{Results}

\subsection*{LChLC assembly of ground-state origamis}

We consider monodisperse B-DNA bundles comprised of 6 double helices crossed-linked in a tight hexagonal arrangement. Such self-assembled filaments may be folded into shapes of programmable twist and curvature through targeted deletions and insertions of base pairs (bp) along each bundle.\cite{Diet09} Following Ref.~\onlinecite{Siav17}, we here focus on four variants of the filaments comprising $15,224$ to $15,240$ nucleotides, with experimentally-determined contour lengths ($l_c$) of \SI{420}{\nano\meter} and bundle diameters ($\sigma$) of \SI{6}{\nano\meter}. A continuum finite-element model based on an elastic rod description of DNA\cite{Kim17} predicts the respective ground states of the different designs to bear negligible (s), \SI{360}{\degree} right-handed (1x-rh), \SI{360}{\degree} left-handed (1x-lh) and \SI{720}{\degree} left-handed (2x-lh) twist about the filament long axis, with negligible net curvature (Fig.~\ref{fig1}).\cite{Siav17}
\par
As a first approximation, we neglect the conformational fluctuations of DNA origamis in solution, and assess the cholesteric arrangement of their respective ground states. To that end, we make use of an efficient and accurate numerical implementation of the Onsager theory extended to the treatment of cholesteric order,\cite{Stra76} which has been extensively discussed elsewhere\cite{Tort17-1,Tort17-2} (see Materials and Methods). In this framework, the reliable investigation of their LChLC assembly requires the input of a mechanical model capable of resolving the local double-helical arrangement of nucleotides within each duplex.\cite{Tomb05} We thus employ the oxDNA model,\cite{Snod15} coarse-grained at the nucleotide level, to represent the origami microscopic structure and interaction potential (Fig.~\ref{fig1}).

\begin{figure*}[]
  \includegraphics[width=\textwidth]{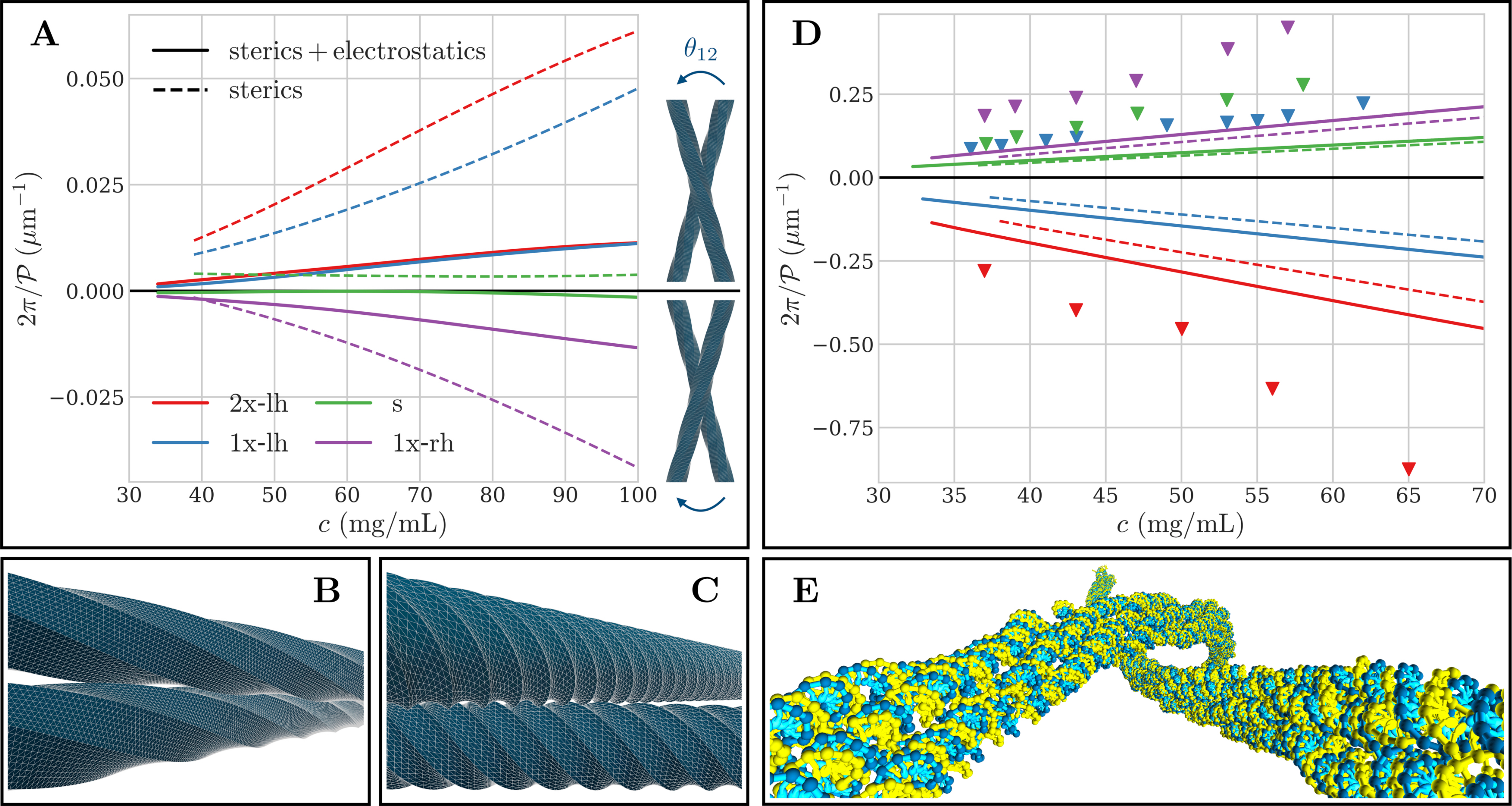}
  \caption{\label{fig2}\textbf{Cholesteric behavior of ground-state and thermalized origamis.} A) Inverse equilibrium cholesteric pitch ($\mathcal{P}$) as a function of particle concentration ($c$) for ground-state filament conformations. Dashed lines denote values obtained by assuming pure steric interactions, and solid lines by accounting for both steric and Debye-H\"uckel repulsion. Positive (resp.~negative) values of $\mathcal{P}$ correspond to LChLC phases bearing right (resp.~left) handedness, as illustrated in the right-hand panel. B) Close-approach configuration of idealized, weakly-twisted right-handed filaments, displaying a left-handed arrangement\cite{Tort17-2} (see Supplementary Movie 1). C) Same as B) for the case of strongly-twisted right-handed filaments, illustrating their entropic preference for right-handed arrangements\cite{Tort17-2} (see Supplementary Movie 2). D) Same as A) for the case of thermalized filaments. Markers denote experimental measurements (from Ref.~\onlinecite{Siav17}). E) Angular configuration minimizing the chiral two-body potential of mean force for thermalized 1x-lh origamis (see Fig.~S1 and Supplementary Section S2), illustrating the predominance of long-wavelength backbone fluctuations over local axial twist in their LChLC assembly.}
\end{figure*}

In the absence of electrostatic interactions, the entropy-induced ordering of ground-state filaments is governed by their axial twist, which is found to stabilize \textit{anti-chiral} LChLC phases --- possessing opposite handedness with respect to the origami twist (Fig.~\ref{fig2}A). This seemingly counterintuitive observation is explained by the fact that the pair excluded volume of weakly-twisted, rod-like filaments is generally minimized by opposite-handed arrangements (Fig.~\ref{fig2}B).\cite{Stra76} Conversely, this entropic preference is reversed in the case of strongly-twisted filaments (Fig.~\ref{fig2}C), which accounts for the weak right-handed phase predicted for the untwisted (s) origamis in terms of the intrinsic right-handed helicity of DNA.\cite{Tomb05} These findings mirror recent results on the LChLC assembly of continuously-threaded particles, for which the quantitative validity of these simple geometric arguments has been investigated in detail.\cite{Tort17-2}
\par
However, these predictions are at odds with the experimental measurements of Ref.~\onlinecite{Siav17}, which instead revealed a general tendency of origami filaments to stabilize \textit{iso-chiral} LChLC phases --- bearing the same handedness as their axial twist. Previous theoretical studies of DNA assemblies have attempted to attribute similar discrepancies to a potential antagonistic influence of electrostatic interactions,\cite{Tomb05,Korn02,Cher08} although the validity of this argument has been disputed by detailed numerical investigations.\cite{Cort17} Here, we instead report that the main effect of the inclusion of longer-ranged Debye-H\"uckel repulsion is to simply unwind the predicted cholesteric pitches by partially screening the chiral nucleotide distribution on the filament surface (see Fig.~S1 and Supplementary Section S2). This finding mirrors the conclusions of Ref.~\onlinecite{Arak01} for the LChLC behavior of bacterial cellulose microcrystals, whose twisted molecular morphologies closely resemble those of the origami ground states, and is consistent with recent all-atom simulations of short B-DNA oligomers, which failed to uncover a statistically-significant chiral contribution attributable to electrostatics in DNA-DNA inter-molecular interactions.\cite{Cort17} Thus, these observations suggest that simple steric and electrostatic repulsion between ground-state filament conformations cannot account for either the handedness or the magnitude of their experimental cholesteric pitches.

\subsection*{Role of conformational statistics}

To assess the influence of conformational statistics on their cholesteric ordering, we make further use of the oxDNA model\cite{Snod15} to probe the detailed thermal fluctuations of the origami filaments. As in Ref.~\onlinecite{Tort18}, we extend our theoretical framework to flexible particles through its combination with the numerical sampling of the filament conformational space by single-origami molecular dynamics (MD) simulations (see Materials and Methods). This hybrid approach, based on the Fynewever-Yethiraj density functional theory,\cite{Fyne98} has been shown to be quantitatively accurate in dilute assemblies of long and stiff persistent chains, for which the effects of many-particle interactions on conformational statistics are limited\cite{Tort18} (see Supplementary Section S1). This description is therefore well-suited for our purposes, given the large persistence length ($l_p$) of the origami structures ($l_p/l_c \gtrsim 5$\cite{Schi13}) and the low packing fractions of their stable LChLC phases.\cite{Siav17} Remarkably, despite its experimental relevance, this regime of large but finite particle rigidity ($\sigma \ll l_c \lesssim l_p$) may not be easily probed by previous theories of cholesteric order, which either neglect the effects of flexibility altogether\cite{Tomb05,Korn02,Tomb06} or focus on semi-flexible polymers in the coil limit ($\sigma \ll l_p \ll l_c$), effectively treated as contiguous collections of rigid chiral segments.\cite{Odij87}
\par
Our results display a surprising phase-handedness inversion compared to the LChLC behavior of the origami ground-states, as well as a considerable tightening of the corresponding equilibrium pitches (Fig.~\ref{fig2}D). The conjunction of these two factors allows for a convincing overall agreement with the experimental measurements of Ref.~\onlinecite{Siav17}, albeit with a slight offset in the crossover value of the origami twist at which the phase handedness inversion occurs. These effects stem from the emergence of long-wavelength helical deformation modes along the backbone of thermalized origamis, which dominate the chiral component of their potential of mean force over the local surface chirality arising from axial twist (Figs.~\ref{fig2}E,~S1).

\begin{figure}[]
  \includegraphics[width=\columnwidth]{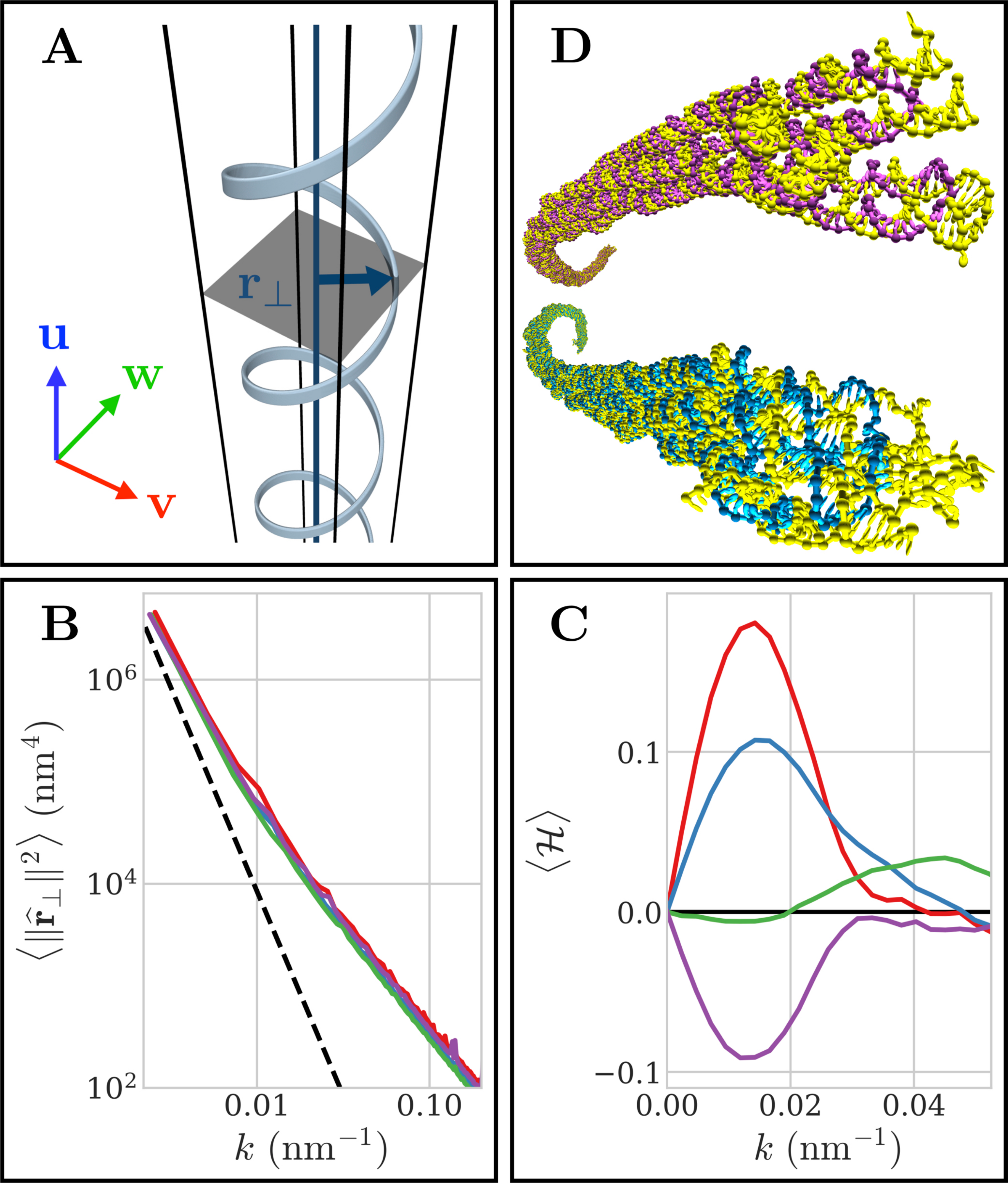}
  \caption{\label{fig3}\textbf{Conformational fluctuations and solenoidal writhe.} A) Transverse deformation vector $\mathbf{r}_\bot$ for an arbitrary backbone conformation. B) Transverse fluctuation spectrum of each origami variant. The dashed line represents the theoretical scaling behavior of generic semi-flexible filaments with bending rigidity $l_p/l_c=8$ at low deformation wavenumbers ($k$) (see Supplementary Section S5). Colors are as in Fig.~\ref{fig2}. C) Net backbone helicity ($\langle\mathcal{H}\rangle$) as a function of $k$. Positive (resp.~negative) values denote a statistical bias towards right-handed (resp.~left-handed) deformation modes. The data is smoothed using a Savitzky-Golay filter of order 9 to facilitate visualization.\cite{Savi64} Colors are as in Fig.~\ref{fig2}. D) Example simulated conformations of 1x-rh (top) and 1x-lh (bottom) origamis, respectively displaying characteristic left- and right-handed backbone helicities.}
\end{figure}

This long-ranged, super-helical (or \textit{solenoidal}) writhe may be quantified by Fourier analysis of the filament backbone conformations (Fig.~\ref{fig3}A, see Materials and Methods). The transverse fluctuation spectra obtained using the oxDNA model are found to be consistent with the asymptotic scaling behavior of persistent chains in the limit of long-wavelength deformations for typical experimental values of the filament bending rigidity\cite{Siav17} (Fig.~\ref{fig3}B). In this regime, the net backbone helicity of each origami variant is found to bear the opposite handedness to the axial twist of its ground state, with left-handed (right-handed) filaments predominantly favoring right-handed (left-handed) helical conformations, respectively (Figs.~\ref{fig3}C-D). Note that the antisymmetric character of the helicity measure $\mathcal{H}(k)$ (see Materials and Methods) imposes that $\mathcal{H}(0)=0$, and leads to the observation of a peak for $\mathcal{H}$ in Fig.~\ref{fig3}C at the smallest accessible deformation wavenumber $k_{\rm min} = 1/l_c$. This peak should therefore be regarded as an effect of the finite size of the filaments, and may \textit{a priori} not be interpreted as evidence of a physical lengthscale indicating a preferred specific helical pitch for the origami shape fluctuations.
\par
The geometric argument of Fig.~\ref{fig2}B, applied to systems of weakly-curled helices, predicts such conformations to display an entropic preference for opposite-handed arrangements.\cite{Tort17-2} In this case, the stabilization of iso-chiral phases of twisted origami filaments therefore arises from their propensity for long-ranged, anti-chiral deformations under the effects of thermal fluctuations. This original chirality amplification mechanism is further evidenced by the relative insensitivity of our results to the inclusion of electrostatic interactions (Fig.~\ref{fig2}D), as the typical lengthscales of the resulting backbone helicities are considerably larger than the experimental Debye screening length ($\lambda_D \simeq \SI{0.6}{\nano\meter}$)\cite{Siav17} (Figs.~\ref{fig3}C-D).

\subsection*{Ground-state structure and helical fluctuations}

The origin of this fluctuation-stabilized solenoidal writhe, and of its dependence on filament twist, lies in the geometric constraints imposed by inter-helical crossovers in the origami design. In the untwisted origami (s), the crossover separation is designed to exactly match the DNA pitch, so that crossovers between adjacent helices are separated by 21 base pairs --- i.e., two full helical turns. In the left-handed designs, for instance, the number of base pairs between crossovers in certain sections of the filaments is reduced by one (known as a ``deletion''). Assuming the origamis to be straight and untwisted, the resulting over-twist of the individual duplexes is given in the ``initial'' column of Table~\ref{tab:writhe}.  The stress arising from this over-twist in the duplexes may be reduced if the origami as a whole adopts a left-twisted configuration about its long axis, as this reduces the net duplex twist between junctions.\cite{Diet09} This redistribution of the stress leads to a decrease in the over-twist in the duplexes of roughly \SI{30}{\percent} upon going from the ``initial'' untwisted origami to the ground state (Fig.~\ref{fig1}), as shown in Table~\ref{tab:writhe}.

\begin{table}
\caption{\textbf{Duplex twist and backbone writhe in thermalized and unthermalized origamis.} The over-twist density $\Delta{\rm Tw} \equiv {\rm Tw}-{\rm Tw}_0$ is averaged over the 6 constituent duplexes of each origami design (see Materials and Methods). The ``initial'' column corresponds to values obtained by assuming that the origamis adopt a straight and untwisted conformation. The writhe density $\rm{Wr}$ of the origami centerlines is smaller than the last significant digit for all ``initial'' and ground-state filaments, and is therefore only reported in the case of the thermalized systems. All thermalized values are further averaged over the full ensemble of simulated conformations for each origami variant.} 
{\begin{tabular}{c c c c c} \toprule
 state & initial & ground & \multicolumn{2}{c}{thermalized} \\ 
  \cmidrule(lr){2-2} \cmidrule(lr){3-3} \cmidrule(lr){4-5} 
   &  $\Delta{\rm Tw}$ & $\Delta{\rm Tw}$ &  $\langle\Delta{\rm Tw}\rangle$& $\langle\rm{Wr}\rangle$ \\
 design  & (${\rm turns}/\SI{}{\micro\meter}$) &(${\rm turns}/\SI{}{\micro\meter}$) &(${\rm turns}/\SI{}{\micro\meter}$) &($\SI{}{\per\milli\meter}$) \\ \midrule
 2x-lh & $15.07$& $10.08$ & $4.05$ & $73.3$ \\ 
 1x-lh & $7.00$ & $4.57$ & $3.20$ & $37.6$ \\ 
 s & $0$ &$ 0$ & $0.61$& $-6.1$ \\ 
 1x-rh & $-6.42$ &$-4.16$ & $-2.72$ &$-38.4$ \\ \bottomrule
\end{tabular}}
\label{tab:writhe}
\end{table}

The residual over-twist of the duplexes is further found to be substantially reduced by the thermal fluctuations. This is achieved by the origamis preferentially adopting writhed configurations with the same sign as the twist stress in the duplexes (see Table~\ref{tab:writhe} and Supplementary Section S6). In other words, when left-handed origamis fluctuate to bear a right-handed helical writhe, the elastic cost of bending is partially offset by a reduction in the residual over-twist of the DNA helices --- while left-handed backbone conformations are energetically penalized by a further over-winding of the duplexes. Conversely, in the case of right-twisted origamis, the required base-pair insertions lead to an under-winding of the individual DNA helices, which in turn favors a left-handed solenoidal writhe.
\par
The observed offset in the filament phase-handedness inversion behavior, apparent in Fig.~\ref{fig2}D, could thus be partially explained in terms of a small misestimate of the equilibrium duplex twist density $\rm{Tw}_0$, as the equilibrium helical pitch of B-DNA within constrained origami structures may slightly differ from the unconfined value $1/\rm{Tw}_0 \simeq \SI{10.5}{\bp}$ assumed in both the computation of the origami ground states\cite{Kim17} and the parametrization of the oxDNA model.\cite{Snod15} Additional possible sources of error include other potential shortcomings of the oxDNA model, such as our use of sequence-averaged mechanics for DNA, or the limitation of soft non-bonded interactions to simple Debye-H\"uckel electrostatics.\cite{Snod15} The overestimations in the magnitude of our cholesteric pitch predictions (Fig.~\ref{fig2}D) are further consistent with the symmetry limitations of the theory, in which long-ranged biaxial correlations arising from broken local cylindrical invariance are neglected.\cite{Tort17-1} The limited extent of these discrepancies, relative to the vast gap between molecular and cholesteric lengthscales, combined with the satisfactory experimental agreement achieved in terms of isotropic/cholesteric binodal concentrations (Table~\ref{tab:coex}) and in the magnitude of the underlying macroscopic curvature elasticities (see Fig.~S2 and Supplementary Section S3), nonetheless evidence the ability of the theory to correctly capture the basic physics of LChLC assembly in our case.

\begin{table}
\caption{\textbf{Isotropic/cholesteric coexistence concentrations for thermalized untwisted origamis.} $c_{\rm st+el}$ and $c_{\rm st}$ denote the theoretical predictions obtained by taking into account steric inter-particle repulsion with and without electrostatic interactions, respectively (see Materials and Methods). Results are compared with the experimental measurements of Ref.~\onlinecite{Siav17}.}
{\begin{tabular}{l c c c} \toprule
 binodal & $c_{\rm st}$ & $c_{\rm st+el}$ & Ref.~\onlinecite{Siav17} \\ \midrule
 isotropic & \SI{31.8}{\gram/\liter} & \SI{28.3}{\gram/\liter} & \SI{28}{\gram/\liter} \\
 cholesteric & \SI{36.7}{\gram/\liter} & \SI{32.2}{\gram/\liter} & \SI{37}{\gram/\liter} \\ \bottomrule
\end{tabular}}
\label{tab:coex}
\end{table}

\section*{Discussion}

We have presented the successful application of an extended Onsager theory to the quantitative description of LChLC order in systems of long DNA origami filaments. Its combination with an accurate conformational sampling scheme demonstrates that phase chirality in this case results from the weak, fluctuation-stabilized solenoidal writhing of the filament backbones, and is therefore largely governed by intra-molecular mechanics. Such long-wavelength, chiral deformation modes, which dominate cholesteric assembly over the much shorter length-scales associated with the twisted morphology of the ground state, are further shown to be linked to the ground-state structure in a non-trivial fashion, as illustrated by the stabilization of anti-chiral deformation modes through twist-writhe conversion of the filament elastic energy.
\par
The net helicity of these backbone fluctuations is found to originate from the weak over- or under-winding of the constituent duplexes in the origami ground states. Similar geometrical frustration phenomena have been shown to widely regulate equilibrium morphology in cohesive bundles of generic chiral filaments,\cite{Hall16} which may be found in the molecular structure of a number of flexible cholesteric mesogens ranging from amyloid fibrils\cite{Adam10} to the protein coat of filamentous viruses.\cite{Grel03} The LChLC assembly of such colloids could therefore be expected to be similarly affected by potential solenoidal deformation modes, thus giving credence to the hypothetical ``corkscrew model'' first proposed in Ref.~\onlinecite{Grel03} to explain the puzzling cholesteric behavior of virus suspensions. This chirality amplification mechanism represents a marked shift from the prevailing theoretical models, in which the macroscopic breaking of mirror symmetry has generally been attributed to the inter-molecular interactions arising from the chiral morphology of the molecular ground state,\cite{Tomb05,Tomb06,deMi16} and more broadly suggests a novel self-assembly paradigm for LChLCs in which subtle, long-ranged conformational features --- rather than local chemical structure --- dictate macroscopic chiral organization.
\par
Finally, the current study, together with the experiments of Ref.~\onlinecite{Siav17}, may provide a new framework to systematically explore the link between molecular properties and supra-molecular organization --- and illustrates how the unique ability of DNA origamis to assemble into programmable shapes of near-arbitrary complexity may be fruitfully combined with the capacity of our theoretical description to rationalize their phase behavior, in order to elucidate the hierarchical self-assembly of complex, chiral macroscopic materials.

\begin{acknowledgments}
JPKD and MMCT gratefully acknowledge \'E.~Grelet and Z.~Dogic for helpful discussions. This project has received funding from the European Union's Horizon 2020 research and innovation programme under the Marie Sk\l{}odowska-Curie Grant Agreement No.~641839. The authors would like to acknowledge the use of the University of Oxford Advanced Research Computing (ARC) facility in carrying out this work (\href{http://dx.doi.org/10.5281/zenodo.22558}{http://dx.doi.org/10.5281/zenodo.22558}). MMCT made use of time on the ARCHER UK National Supercomputing Service (\href{http://www.archer.ac.uk}{http://www.archer.ac.uk}) granted via the UK High-End Computing Consortium for Biomolecular Simulation, HECBioSim (\href{http://www.hecbiosim.ac.uk}{http://www.hecbiosim.ac.uk}). MMCT is grateful to the UK Materials and Molecular Modelling Hub for computational resources, which is partially funded by EPSRC (EP/P020194/1). GM is grateful for financial support from the Department of Science and Technology India in the form of an INSPIRE faculty grant (DST/INSPIRE/04/2014/002085), and acknowledges the computational facilities provided by Calcul Qu\'ebec (\href{www.calculquebec.ca}{www.calculquebec.ca}) and Compute Canada (\href{www.computecanada.ca}{www.computecanada.ca}). DP is supported by the EPRSC Centre for Doctoral Training in Theory and Modelling in the Chemical Sciences (EP/L015722/1).
\end{acknowledgments}

\section*{Author Contributions}
\noindent
MMCT developed the theory, conducted its implementation, carried out the numerical calculations and wrote the manuscript. GM and DP performed the origami simulations. JPKD and MMCT devised the study, analyzed the results and proofread the completed manuscript. 

\section*{Data availability}
\noindent
The numerical code employed for all density-functional and related calculations may be found at \href{https://github.com/mtortora/chiralDFT}{https://github.com/mtortora/chiralDFT}. The oxDNA simulation package is also available online (\href{https://dna.physics.ox.ac.uk}{https://dna.physics.ox.ac.uk}). Input files will be provided upon request to the authors.

\section*{Materials and Methods} \label{sec:Methods}

\subsection*{MD simulations setup} \label{subsec:MD}
Single-origami simulations were run for each of the 6-helix-bundle designs in Ref.~\onlinecite{Siav17} using the oxDNA coarse-grained model, which represents DNA as a collection of rigid nucleotides interacting through excluded volume, Debye-H\"uckel, stacking, hydrogen- and covalent-bonding potentials.\cite{Snod15} Calculations were performed on GPUs in the canonical ensemble using an Andersen-like thermostat and sequence-averaged DNA thermodynamics, assuming room-temperature conditions ($T=\SI{293}{\kelvin}$) and fixed monovalent salt concentration $c_{{\rm Na}^+} \! = \SI{0.5}{\molar}$. This value was chosen in slight excess of the experimental salt concentration $c_{{\rm Na}^+} \! = \SI{0.26}{\molar}$,\cite{Siav17} employed throughout the rest of the paper, in order to limit computational costs. The effects of this approximation on origami conformational statistics are expected to be minimal in the context of the simplified oxDNA treatment of electrostatics.\cite{Snod15} Relaxation was achieved through equilibration runs of $\mathcal{O}(10^6)$ MD steps starting from the origami ground state, and production runs of $\mathcal{O}(10^9)$ steps were conducted to generate $\mathcal{O}(10^3)$ uncorrelated conformations for each origami variant. The statistical independence of the resulting conformations was assessed by ensuring the vanishing autocorrelation of their end-to-end separation distance. 

\subsection*{Conformational analysis} \label{subsec:analysis}
The discretized origami backbones are obtained by averaging the center-of-mass locations of their bonded nucleotides over the 6 constituent duplexes within each transverse plane along the origami contour.\cite{Diet09} We define the molecular frame $\mathcal{R}=\big [\mathbf{u} \; \mathbf{v} \; \mathbf{w} \big]$ of each conformation as the principal frame of its backbone gyration tensor, such that $\mathbf{u}$ and $\mathbf{v}$ correspond to the respective direction of maximum and minimum dispersion of the origami backbone.\cite{Tort18} Shape fluctuations are described by the contour variations of the transverse position vector, 
\begin{equation}
\mathbf{r}_\bot (s) = \mathbf{r}(s)-r_u(s) \mathbf{u}, 
\end{equation}
with $\mathbf{r}(s)$ the position of the discretized backbone segment with curvilinear abscissa $s$ and $r_u(s) \equiv \mathbf{r}(s)\cdot \mathbf{u}$, assuming the backbone center of mass to be set to the origin of the frame. Denoting by $\Delta s$ the curvilinear length of each segment, the Fourier components of $\mathbf{r}_\bot$ read as 
\begin{equation}
  \widehat{\mathbf{r}}_\bot (k) = \sum_s \Delta s \, \mathbf{r}_\bot (s) \times e^{-2i\pi ks}.
\end{equation}
Using the convolution theorem, the spectral coherence between the two transverse components of an arbitrary backbone deformation mode may be quantified by their Fourier-transformed cross-correlation function $\widehat{c}_{vw}$,
\begin{equation}
  \widehat{c}_{vw}(k) = \widehat{r}_{\bot v}(k) \times \widehat{r}^{\:*}_{\bot w}(k),
\end{equation}
where $\widehat{r}_{\bot x} = \widehat{\mathbf{r}}_\bot \cdot \mathbf{x}$ for $\mathbf{x}\in\{\mathbf{v},\mathbf{w}\}$ and $\widehat{r}_{\bot w}^{\:*}$ is the complex conjugate of $\widehat{r}_{\bot w}$. It is shown in Supplementary Section S4 that an helicity order parameter $\mathcal{H}(k)$ for a deformation mode with arbitrary wavenumber $k$ about the filament long axis $\mathbf{u}$ may be derived in the form
\begin{equation}
  \label{eq:h_corr}
  \mathcal{H}(k) = \frac{2 \times \Im \big \{ \widehat{c}_{vw}(k) \big \} }{\widehat{c}_{vv}(k) + \widehat{c}_{ww}(k)},
\end{equation}
with $\Im\{ \widehat{c}_{vw} \big \}$ the imaginary part of $\widehat{c}_{vw}$. One may check that $-1 \leq \mathcal{H}(k) \leq 1$, with $\mathcal{H}(k) = \pm 1$ if and only if the two transverse Fourier components bear equal amplitudes and lie in perfect phase quadrature. In this case, $\widehat{\mathbf{r}}_\bot (k)$ describes an ideal circular helical deformation mode with pitch $1/k$ and handedness determined by the sign of $\mathcal{H}$.

\subsection*{Determination of duplex twist and backbone writhe} \label{subsec:twist-writhe}

Let $\mathbf{r}_{i,1}(s_i)$ and $\mathbf{r}_{i,2}(s_i)$ be a set of continuous curves interpolating the positions of the nucleotide centers of mass of the $i$-th constituent duplex of an arbitrary origami conformation. The unit tangent and normal vectors $\mathbf{t}_i$ and $\mathbf{n}_i$ at a given curvilinear abscissa $s_i$ respectively read as
\begin{gather*}
  \mathbf{t}_i(s_i) \equiv \frac{d\mathbf{r}_i(s_i)}{ds_i}, \\
  \mathbf{n}_i(s_i) \equiv \frac{\mathbf{r}_{i,1}(s_i)-\mathbf{r}_{i,2}(s_i)}{\lVert \mathbf{r}_{i,1}(s_i)-\mathbf{r}_{i,2}(s_i) \rVert},
\end{gather*}
where $\mathbf{r}_i\equiv (\mathbf{r}_{i,1}+\mathbf{r}_{i,2})/2$ is the continuous duplex centerline with contour length $l_i$. The average twist density $ {\rm Tw}$ of each individual duplex may then be obtained from the sum of the local stacking angles between consecutive base pairs,\cite{Saya10}
\begin{equation}
  \label{eq:twist}
  {\rm Tw}_i = \frac{1}{2\pi l_i} \int_0^{l_i} ds_i \, \mathbf{t}_i(s_i)\cdot \bigg\{\mathbf{n}_i(s_i)\times\frac{d\mathbf{n}_i(s_i)}{ds_i}\bigg\}.
\end{equation}
For stiff origamis, whose centerline curve $\mathbf{r} \equiv \sum_{i=1}^{6} \mathbf{r}_i /6$ does not display any turning points in $\mathbf{u}$ ($dr_u/ds > 0$), the so-called \textit{polar writhe} $ {\rm Wr}$ of the filament backbone simply reduces to the local contribution~\cite{Berg06}
\begin{equation}
  \label{eq:writhe}
  {\rm Wr} = \frac{1}{2\pi l_c}  \int_0^{l_c} ds \, \frac{\mathbf{u}\cdot \Big\{\mathbf{t}(s)\times\frac{d\mathbf{t}(s)}{ds}\Big\}}{1+\mathbf{u}\cdot \mathbf{t}(s)},
\end{equation}
where $\mathbf{t} \equiv d\mathbf{r}/ds$ is the unit backbone tangent vector. It may then be shown that  ${\rm Wr} > 0$ (resp.~${\rm Wr} < 0$) if $\mathbf{t}$ winds about $\mathbf{u}$ in a right-handed (resp.~left-handed) fashion.\cite{Berg06} Eqs.~\eqref{eq:twist} and~\eqref{eq:writhe} are evaluated numerically through standard quadrature methods, using cubic spline interpolations for all discrete curves.\cite{Saya10} A Savitzky-Golay filter of order 9\cite{Savi64} was preliminarily applied to the backbone curve to weed out irrelevant short-wavelength contour fluctuations arising from our geometric definition of the origami centerline.\cite{Saya10} The last $\mathcal{O}(10)$ base-pair planes at each of the filament extremities were excluded from the calculations to limit the influence of end effects.

\subsection*{Molecular theory of cholesteric order} \label{subsec:DFT}
We consider a cholesteric phase of director field $\mathbf{n}$ and helical axis $\mathbf{e}_z$ in the laboratory frame $\mathcal{R}_{\rm lab} \equiv \big [\mathbf{e}_x \; \mathbf{e}_y\; \mathbf{e}_z\big]$, whose continuum Helmholtz free energy density is expressed by the Oseen-Frank functional,\cite{deGe93}
\begin{equation}
  \label{eq:Frank}
 \mathpzc{f} = \mathpzc{f}_0 + \frac{1}{2} \Big\{ K_2 \left(\mathbf{n} \cdot \left[\nabla\times\mathbf{n}\right]\right)^2 + 2k_t\left(\mathbf{n} \cdot \left[\nabla\times\mathbf{n}\right]\right) \Big\}.
\end{equation}
Given the high anisotropy of the origami structures and the low packing fractions marking the onset of their LChLC organization,\cite{Siav17} the mean-field free energy $\mathpzc{f}_0$ of their reference nematic state with uniform director $\mathbf{n}\equiv\mathbf{e}_x$ may be written in a generalized Onsager form, based on the second-virial kernel $\kappa$\cite{Fyne98} (see Supplementary Section S1),
\begin{multline}
  \label{eq:exc}
 \kappa(\theta,\theta') = \int d\mathbf{r}_{12}  \oiint d \mathcal{R}_1 d\mathcal{R}_2 \, \overline{f}(\mathbf{r}_{12}, \mathcal{R}_1, \mathcal{R}_2)\\
  \times \delta(\cos\theta_1-\cos\theta)\delta(\cos\theta_2-\cos\theta'),
\end{multline}
with $\delta$ the Dirac distribution and $\overline{f}$ the Mayer $f$-function averaged over all pairs of accessible molecular conformations,
\begin{equation}
  \label{eq:f_ave}
  \overline{f}(\mathbf{r}_{12}, \mathcal{R}_1, \mathcal{R}_2) = \Big\langle \Big \langle e^{ -\beta U_{\rm inter}(\mathbf{r}_{12}, \mathcal{R}_1, \mathcal{R}_2)} -1 \Big \rangle \Big\rangle.
\end{equation}
In Eq.~\eqref{eq:f_ave}, $U_{\rm inter}(\mathbf{r}_{12}, \mathcal{R}_1, \mathcal{R}_2)$ denotes the inter-molecular interaction energy of two arbitrary origami conformations with center-of-mass separation $\mathbf{r}_{12}$ and respective molecular-frame orientations $\mathcal{R}_{1,2}$, and $\langle\cdot\rangle$ is the ensemble average over the single-origami conformations generated by MD simulations.\cite{Tort18} Local uniaxial order is described by the equilibrium orientation distribution function $\psi(\cos\theta) \equiv \psi(\mathbf{e}_x\cdot\mathbf{u})$, quantifying the dispersion of the origami long axes $\mathbf{u} = \mathcal{R}\cdot\mathbf{e}_x$ about $\mathbf{e}_x$. $\psi$ is obtained by functional minimization of $\mathpzc{f}_0$ at fixed number density $\rho$ and inverse temperature $\beta=1/k_b T$,\cite{Tort17-1}
\begin{equation}
  \label{eq:psi}
  \psi (\cos\theta) = \frac{1}{Z} \exp\bigg\{\frac{\rho}{4\pi^2} \int_{-1}^1 d\cos\theta' \, \psi(\cos\theta') \kappa(\theta, \theta') \bigg\},
\end{equation}
with $Z$ a Lagrange multiplier ensuring the normalization of $\psi$. The Oseen-Frank twist elastic modulus $K_2$ and chiral strength $k_t$ read as (see Supplementary Section S1)
\begin{align}
  \label{eq:K2}
  \beta K_2 &= \frac{\rho^2}{2} \int_V d\mathbf{r}_{12} \oiint d\mathcal{R}_1 d\mathcal{R}_2 \,  \overline{f}(\mathbf{r}_{12}, \mathcal{R}_1, \mathcal{R}_2) \\ \nonumber &\qquad \times \dot{\psi}(\cos\theta_1) \dot{\psi}(\cos\theta_2) r_z^2 u_{1y} u_{2y}, \\[+1mm]
  \label{eq:kt}
  \beta k_t &= \frac{\rho^2}{2} \int_V d\mathbf{r}_{12} \oiint d\mathcal{R}_1 d\mathcal{R}_2 \,  \overline{f}(\mathbf{r}_{12}, \mathcal{R}_1, \mathcal{R}_2) \\ \nonumber &\qquad \times \psi(\cos\theta_1) \dot{\psi}(\cos\theta_2) r_z u_{2y},
\end{align}
with $r_z = \mathbf{r}_{12} \cdot \mathbf{e}_z$, $u_{iy} = \mathbf{u}_i \cdot \mathbf{e}_y$ and $\dot{\psi}$ the first derivative of $\psi$. The equilibrium cholesteric pitch is determined by the competition between chiral torque and curvature elasticity, and is obtained by minimization of the elastic contribution to the free energy density $\mathpzc{f}$ (Eq.~\eqref{eq:Frank}),\cite{Stra76}
\begin{equation}
  \label{eq:pitch}
  \mathcal{P} = 2\pi \frac{K_2}{k_t}.
\end{equation}
Eqs.~\eqref{eq:exc},~\eqref{eq:K2} and~\eqref{eq:kt} are evaluated through optimized virial integration techniques\cite{Tort17-2} over 16 independent runs of $10^{13}$ Monte-Carlo (MC) steps, using oxDNA-parametrized Debye-H\"uckel and steric inter-nucleotide repulsion for the inter-molecular potential $U_{\rm inter}$.\cite{Snod15} The conformational average in Eq.~\eqref{eq:f_ave} is performed by stochastic sampling over the simulated origami conformations in Eqs.~\eqref{eq:exc},~\eqref{eq:K2} and~\eqref{eq:kt}.\cite{Tort18} Eq.~\eqref{eq:psi} is solved through standard numerical means.\cite{Herz84} Convergence was ensured by verifying the numerical dispersion of the computed pitches (Eq.~\eqref{eq:pitch}) to be less than $\SI{10}{\percent}$ across the results of the 16 MC runs, using independent bootstrap samples of the ensemble of simulated conformations. Binodal points were calculated by equating chemical potentials and osmotic pressures in the isotropic and cholesteric phase, and solving the resulting coupled coexistence equations numerically.\cite{Tort17-1} Mass concentrations were obtained assuming a molar weight of $\SI{650}{\dalton}$ per base pair.


\bibliography{refs}
\bibliographystyle{Science}

\end{document}


\title{Supplementary information: Chiral shape fluctuations and the origin of chirality in cholesteric phases of DNA origamis}
\date{\today}
\author{Maxime M.C. Tortora}
\thanks{Current address: Laboratory of Biology and Modeling of the Cell, \'Ecole Normale Sup\'erieure de Lyon, 46, all\'ee d’Italie, 69364 Lyon Cedex 07, France}
\thanks{correspondence to \href{mailto:maxime.tortora@ens-lyon.fr}{maxime.tortora@ens-lyon.fr}}
\affiliation{Physical and Theoretical Chemistry Laboratory, Department of Chemistry, University of Oxford, South Parks Road, Oxford OX1 3QZ, United Kingdom}
\author{Garima Mishra}
\affiliation{Department of Physics, Indian Institute of Technology Kanpur, Kanpur 208016, India\looseness=-1}
\author{Domen Pre\v{s}ern}
\affiliation{Physical and Theoretical Chemistry Laboratory, Department of Chemistry, University of Oxford, South Parks Road, Oxford OX1 3QZ, United Kingdom}
\author{Jonathan P.K. Doye}
\affiliation{Physical and Theoretical Chemistry Laboratory, Department of Chemistry, University of Oxford, South Parks Road, Oxford OX1 3QZ, United Kingdom}

\maketitle

\section{Fynewever-Yethiraj density functional theory for LChLCs} \label{app:FY}

In the context of classical density functional theory, the Helmholtz free energy of a system of polyatomic molecules may be written in the general form\cite{Chan86-1}
\begin{equation}
  \label{eq:free_energy}
  \mathscr{F}[\rho_m] = \mathscr{F}_{\rm id}[\rho_m] + \mathscr{F}_{\rm ex}[\rho_m],
\end{equation}
where the microscopic density $\rho_m$ generally depends on the discrete set of atom positions $\{\mathbf{r}_i\}_{i\geq 1}$ and bond orientations $\{\mathcal{R}_j\}_{j\geq 1}$ characterizing the full microscopic state of each individual constituent particle. The center-of-mass position $\mathbf{r}$ and molecular orientation $\mathcal{R}$ of a given particle in any conformation are uniquely determined by the specification of all internal degrees of freedom $\{\mathbf{r}_i\}$ and $\{\mathcal{R}_j\}$, so that one may write, without loss of generality,
\begin{equation}
  \rho_m\big(\{\mathbf{r}_i\},\{\mathcal{R}_j\}\big) = \rho_m\big (\mathbf{r}, \mathcal{R}, \{\mathcal{X}\}\big), 
\end{equation}
with $\{\mathcal{X}\} \equiv \big(\{\mathbf{r}_i\}_{i \geq 2}, \{\mathcal{R}_j\}_{j \geq 2}\big)$. Let $\mathbf{r}'_i$ and $\mathcal{R}'_j$ be the respective projections of $\mathbf{r}_i$ and $\mathcal{R}_j$ in the molecular frame $\mathcal{R}$ centered on $\mathbf{r}$,
\begin{align}
  \label{eq:rp}
  \mathbf{r}'_i & \equiv  \mathcal{R}^{\sf T} \cdot (\mathbf{r}_i-\mathbf{r}),\\
  \label{eq:Rp}
  \mathcal{R}'_j &\equiv \mathcal{R}^{\sf T} \cdot \mathcal{R}_j,
\end{align}
with $ \mathcal{R}^{\sf T}$ the matrix transpose of $\mathcal{R}$. The Fynewever-Yethiraj (FY) approximation postulates that $\rho_m$ may be cast in the decoupled form\cite{Fyne98,vanW12,Tort18}
\begin{equation}
  \label{eq:fy_dens}
   \rho_m\big (\mathbf{r}, \mathcal{R}, \{\mathcal{X}\}\big) \simeq \rho(\mathbf{r},\mathcal{R}) \times P\big(\{\mathcal{X}'\}\big),
\end{equation}
where $\{\mathcal{X}'\} \equiv \big(\{\mathbf{r}'_i\}_{i \geq 2}, \{\mathcal{R}'_j\}_{j \geq 2}\big)$. In Eq.~\eqref{eq:fy_dens}, $\rho$ corresponds to the \textit{molecular density} describing the global distribution of particle centers of mass $\mathbf{r}$ and orientations $\mathcal{R}$ throughout the sample, while $P$ quantifies the distribution of the conformational degrees of freedom $\mathbf{r}'_i$ and $\mathcal{R}'_j$ in the local molecular frame, subject to the respective normalization constraints\cite{vanW12}
\begin{align}
  \label{eq:norm_P}
  \int d\{\mathcal{X}'\} \,  P\big(\{\mathcal{X}'\}\big) &= 1, \\
  \label{eq:norm_rho}
    \int_V d\mathbf{r}\oint d\mathcal{R} \, \rho(\mathbf{r},\mathcal{R}) &= N.
\end{align}
In the FY theory, $P$ is assumed to be entirely determined by the intra-molecular interaction potential $U_{\rm intra} = U_{\rm intra}\big(\{\mathcal{X}'\}\big)$, so as to be independent of the overall position $\mathbf{r}$ and orientation $\mathcal{R}$ of the molecule. In the absence of external fields, this approximation amounts to neglecting the effects of many-particle interactions on conformational statistics, and is therefore only rigorously justifiable in the case of highly-stiff molecules, for which the accessible conformational space is largely independent of density in the regime of low-to-moderate particle packing fractions.\cite{Tort18} 
\par
Discarding the effects of inter-molecular interactions, the \textit{ideal} component $\mathscr{F}_{\rm id}$ of the Helmholtz free energy functional is given by\cite{Jaff01,vanW12}
\begin{multline*}
  \beta\mathscr{F}_{\rm id}[\rho_m] = \int_V d\mathbf{r} \oint d\mathcal{R} \int d\{\mathcal{X}\} \, \rho_m\big (\mathbf{r}, \mathcal{R}, \{\mathcal{X}\}\big)  \\ \times \Big\{ \log \big[\lambda_{\rm dB}^3 \rho_m\big (\mathbf{r}, \mathcal{R}, \{\mathcal{X}\}\big)\big] -1 + \beta U_{\rm intra}\big(\{\mathcal{X}\}\big)\Big\},
\end{multline*}
with $\lambda_{\rm dB}$ the thermal de Broglie wavelength. Using Eqs.~\eqref{eq:fy_dens}--\eqref{eq:norm_rho},
\begin{equation}
   \label{eq:f_id_1}
  \beta\mathscr{F}_{\rm id}[\rho] = \int_V d\mathbf{r} \oint d\mathcal{R} \,\rho(\mathbf{r},\mathcal{R}) \big\{\log \big[\lambda^3 \rho (\mathbf{r}, \mathcal{R})\big] -1 \big\},
\end{equation}
where the lengthscale $\lambda$ now reads as 
\begin{multline*}
  \lambda = \lambda_{\rm dB} \exp\bigg\{ \frac{1}{3} \int d\{\mathcal{X}'\}\,P\big(\{\mathcal{X}'\}\big) \\ \times \Big[ \log P\big(\{\mathcal{X}'\}\big) +\beta U_{\rm intra}\big(\{\mathcal{X}'\}\big) \Big] \bigg\},
\end{multline*}
in which we used the change of variables of Eqs.~\eqref{eq:rp} and~\eqref{eq:Rp}, with unit Jacobian determinant. Note that $\lambda$ generally depends on intra-molecular properties as well as temperature, but is independent of $\rho$. In the case of a prolate nematic phase with arbitrary director field $\mathbf{n}(\mathbf{r})$, the molecular density function $\rho$ takes the form
\begin{equation}
  \label{eq:rho_nem_a}
  \rho(\mathbf{r},\mathcal{R}) = \overline{\rho} \psi\big\{ \mathbf{u} \cdot \mathbf{n}(\mathbf{r}) \big\},
\end{equation}
where $\overline{\rho} \equiv N/V$ is the molecular number density, and the orientation distribution function (ODF) $\psi$ describes the ordering of the long molecular axes $\mathbf{u}\equiv \mathcal{R}\cdot \mathbf{e}_x$ about the local director $\mathbf{n}(\mathbf{r})$. Note that Eq.~\eqref{eq:rho_nem_a} is only valid in the limit where the spatial fluctuations of $\mathbf{n}$ are negligible at the molecular lengthscale, as is typical in experimental cholesterics, and in the absence of long-ranged biaxial correlations, as is commonly presumed in theoretical studies.\cite{Wens11,Frez14,Duss15,deMi16,Tort17-1} Eq.~\eqref{eq:rho_nem_a} further assumes the local molecular density $\overline{\rho}$ to be unaffected by director fluctuations, which is expected to be appropriate in the case of the twist deformations characteristic of LChLCs.\cite{Meye82} Let us define the unit-Jacobian transformation 
\begin{equation}
  \label{eq:t_chg}
  \mathcal{R}' \equiv \mathcal{T}(\mathbf{r})^{\sf T} \cdot \mathcal{R},
\end{equation}
with $\mathcal{T}(\mathbf{r})$ a rotation matrix such that
\begin{equation}
  \label{eq:t_transf}
  \mathbf{n}(\mathbf{r}) = \mathcal{T}(\mathbf{r}) \cdot \mathbf{n}(\mathbf{0}) \equiv  \mathcal{T}(\mathbf{r}) \cdot \mathbf{n}_0.
\end{equation}
Eqs.~\eqref{eq:f_id_1} and~\eqref{eq:rho_nem_a} immediately yield
\begin{equation}
  \label{eq:f_id_2}
  \frac{\beta\mathscr{F}_{\rm id}[\psi]}{V} =  4\pi^2  \rho \int_{-1}^1 du'_x \, \psi(u'_x)  \Big \{ \log \big[\rho \lambda^3 \psi(u'_x)\big] -1\Big\},
\end{equation}
where $u'_x \equiv \mathbf{n}_0^{\sf T} \cdot \mathcal{R}' \cdot \mathbf{e}_x$, and we dropped the overline notation from $\overline{\rho}$. Thus, Eq.~\eqref{eq:f_id_2} indicates that $\mathscr{F}_{\rm id}$ is independent of the configuration of the director field $\mathbf{n}$.
\par
In the case of highly stiff and elongated molecules, the \textit{excess} component $\mathscr{F}_{\rm ex}$ of the Helmholtz free energy may be related to the inter-molecular interaction potential $U_{\rm inter}$ at the second virial level through the Onsager mean-field functional,\cite{Jaff01}
\begin{equation}
  \label{eq:f_ex_1}
  \beta \mathscr{F}_{\rm ex} [\rho_m] = -\frac{1}{2} \iint d\mathbf{1} d\mathbf{2} \, \rho_m(\mathbf{1}) \rho_m(\mathbf{2}) f(\mathbf{1},\mathbf{2}),
\end{equation}
where the shorthand $\mathbf{i} \equiv \big(\mathbf{r}_i,\mathcal{R}_i, \{\mathcal{X}_i\} \big)$ refers to the full microscopic degrees of freedom associated with particle $i$, and $f$ is the so-called Mayer function,
\begin{equation*}
  f(\mathbf{1},\mathbf{2})\equiv \exp\big\{-\beta U_{\rm inter} (\mathbf{1},\mathbf{2})\big\} -1.
\end{equation*}
Using Eqs.~\eqref{eq:fy_dens} and~\eqref{eq:rho_nem_a}, Eq.~\eqref{eq:f_ex_1} may be recast as\cite{Tort17-1,Tort18,vanW12}
\begin{multline}
  \label{eq:f_ex_2}
  \beta \mathscr{F}_{\rm ex} [\psi] = -\frac{\rho^2}{2} \iint_V d\mathbf{r}_1 d\mathbf{r}_2 \oiint d\mathcal{R}_1 d\mathcal{R}_2  \\ \times \psi\big\{ \mathbf{u}_1 \cdot \mathbf{n}(\mathbf{r}_1) \big\} \psi\big\{ \mathbf{u}_2 \cdot \mathbf{n}(\mathbf{r}_2) \big\}\overline{f}(\mathbf{r}_1,\mathbf{r}_2,\mathcal{R}_1,\mathcal{R}_2),
 \end{multline}
 with $\overline{f}$ the \textit{conformational average} of the Mayer function,
 \begin{multline}
   \label{eq:f_ma_ave}
   \overline{f}(\mathbf{r}_1,\mathbf{r}_2,\mathcal{R}_1,\mathcal{R}_2) \equiv \iint d\{\mathcal{X}'_1\} d\{\mathcal{X}'_2\} \, P\big(\{\mathcal{X}'_1\}\big) P\big(\{\mathcal{X}'_2\}\big)  \\ \times f(\mathbf{1},\mathbf{2}).
 \end{multline}
 \par
 Note that the integrand in Eq.~\eqref{eq:f_ex_2} is non-zero only if $\overline{f}(\mathbf{r}_1,\mathbf{r}_2,\mathcal{R}_1,\mathcal{R}_2) \neq 0$, i.e., if there exists two molecular conformations with respective center-of-mass positions $\mathbf{r}_{1,2}$ and overall orientations $\mathcal{R}_{1,2}$ such that $U_{\rm inter} \neq 0$. It follows that in the case of short-range interaction potentials, a pair of molecules $1$ and $2$ may physically contribute to the integral in Eq.~\eqref{eq:f_ex_2} only if their center-of-mass separation distance $\mathbf{r}_{12} \equiv \mathbf{r}_2-\mathbf{r}_1$ is of the order of the typical molecular dimensions. Let us introduce the particle barycenter $\mathbf{R} = (\mathbf{r}_1+\mathbf{r}_2)/2$, 
 \begin{equation}
   \label{eq:bary}
   \mathbf{r}_{1,2} = \mathbf{R} \mp \frac{\mathbf{r}_{12}}{2}.
 \end{equation}
 Under the assumptions of Eq.~\eqref{eq:rho_nem_a}, we may thus write\cite{Wens11}
 \begin{equation*}
    \mathbf{n}(\mathbf{r}_i) \simeq  \mathbf{n}(\mathbf{R}) \mp \frac{\nabla\mathbf{n}(\mathbf{R}) \cdot \mathbf{r}_{12}}{2}+ \frac{\nabla^2\mathbf{n}(\mathbf{R}) \mathbin{:} (\mathbf{r}_{12}\otimes\mathbf{r}_{12})}{8}.
 \end{equation*}
with $\otimes$ and $\mathbin{:}$ the respective tensor and double dot products. In the case of a cholesteric phase of axis $\mathbf{e}_z$ and inverse pitch $q\equiv 2\pi/\mathcal{P}$, the helical modulation of the director field takes the form
 \begin{equation*}
   \mathbf{n}(\mathbf{R}) = \cos (qR_z) \,\mathbf{e}_x + \sin (qR_z) \, \mathbf{e}_y,
 \end{equation*}
 where $R_z \equiv \mathbf{R}\cdot \mathbf{e}_z$ and we have chosen the laboratory frame such that $\mathbf{e}_x \equiv \mathbf{n}_0$. Let $\mathcal{T}(\mathbf{R}) \equiv  \big [\mathbf{e}'_x \; \mathbf{e}'_y\; \mathbf{e}'_z\big]$ be a local rotating frame satisfying Eq.~\eqref{eq:t_transf},
 \begin{align*}
  \mathcal{T}(\mathbf{R}) &\equiv
\begin{bmatrix}
\cos(qR_z) & -\sin(qR_z) & 0 \\
\sin(qR_z) & \cos(qR_z) &0 \\
0 & 0 & 1
\end{bmatrix}.
\end{align*}
It is straightforward to show that 
\begin{gather*}
  \nabla\mathbf{n}(\mathbf{R}) = q \, \mathbf{e}'_y \otimes \mathbf{e}'_z, \\
  \nabla^2\mathbf{n}(\mathbf{R}) = - q^2 \,\mathbf{e}'_x \otimes \mathbf{e}'_z \otimes \mathbf{e}'_z,
\end{gather*}
which directly lead to
 \begin{multline}
   \label{eq:psi_exp}
  \psi\big\{ \mathbf{u}_i \cdot \mathbf{n}(\mathbf{r}_i)\big\} =   \psi (u'_{ix}) \mp \frac{q u'_{iy} r'_z}{2}\dot{\psi}( u'_{ix}) \\ - \frac{u'_{ix}}{2} \bigg\{\frac{q r'_z}{2}\bigg\}^2\dot{\psi}( u'_{ix}) + \frac{1}{2} \bigg\{\frac{qu'_{iy} r'_z}{2}\bigg\}^2 \ddot{\psi}(u'_{ix}) + \mathcal{O}(q^3),
 \end{multline}
 where primed quantities are expressed in the rotating frame $\mathcal{T}(\mathbf{R})$, with $u'_{ij} \equiv \mathbf{u}_i \cdot \mathbf{e}'_j$ and $r'_z \equiv \mathbf{r}_{12} \cdot \mathbf{e}'_z$. Plugging Eq.~\eqref{eq:psi_exp} into Eq.~\eqref{eq:f_ex_2}, and using the changes of variables of Eqs.~\eqref{eq:t_chg} and~\eqref{eq:bary}, we obtain
\begin{equation*}
  \mathscr{F} = \int_V d\mathbf{R} \, \mathpzc{f}(\mathbf{R}) \equiv \int_V d\mathbf{R} \, (\mathpzc{f}_0 + \mathpzc{f}_d) ,
\end{equation*}
in which $\mathpzc{f}_0$ is the free energy density of the uniform nematic state with director $\mathbf{n}_0 \equiv \mathbf{e}_x$,
  \begin{multline}
   \label{eq:f0}
   \beta \mathpzc{f}_0[\psi] =  4\pi^2 \rho \int_{-1}^1 du_x \, \psi(u_x)  \Big \{ \log \big[\rho \lambda^3 \psi(u_x)\big] -1\Big\} \\-\frac{\rho^2}{2} \int_V d\mathbf{r}_{12} \oiint d\mathcal{R}_ 1 d\mathcal{R}_2 \, \psi(u_{1x})\psi(u_{2x}) \overline{f}(\mathbf{r}_{12},\mathcal{R}_1,\mathcal{R}_2),
  \end{multline}
where we used Eqs.~\eqref{eq:free_energy} and~\eqref{eq:f_id_2}, and dropped the prime notation from the dummy integration variables. The integration by parts of the second-order terms in Eq.~\eqref{eq:psi_exp} with respect to $\mathbf{R}$ yields the distortion free energy density $\mathpzc{f}_d$ in the form, to quadratic order in $q$,\cite{Stra76} 
\begin{equation}
  \label{eq:f_el}
  \mathpzc{f}_d[\psi] = -k_t [\psi] q + K_2[\psi] \frac{q^2}{2},
\end{equation}
which by term-to-term comparison with the Oseen-Frank free energy (Eq.~(7) in the main text) leads to the microscopic expressions of the chiral strength $k_t$ and twist elastic modulus $K_2$ as direct generalizations of the Poniewierski-Stecki formulae,\cite{Poni79}
\begin{align}
  \label{eq:K2}
  \beta K_2[\psi] &= \frac{\rho^2}{2} \int_V d\mathbf{r}_{12} \oiint d\mathcal{R}_1 d\mathcal{R}_2 \, \overline{f}(\mathbf{r}_{12},\mathcal{R}_1,\mathcal{R}_2) \\ \nonumber &\qquad \times \dot{\psi}(u_{1x}) \dot{\psi}(u_{2x}) u_{1y} u_{2y} r_z^2, \\[+1mm]
    \label{eq:kt}
  \beta k_t[\psi] &= \frac{\rho^2}{2} \int_V d\mathbf{r}_{12} \oiint d\mathcal{R}_1 d\mathcal{R}_2 \, \overline{f}(\mathbf{r}_{12},\mathcal{R}_1,\mathcal{R}_2) \\ \nonumber &\qquad \times \psi(u_{1x}) \dot{\psi}(u_{2x}) u_{2y} r_z,
\end{align}
from which one recovers Eqs.~(11) and~(12) of the main text, with $\cos\theta_i \equiv u_{ix}$.
\par
In the limit of long-wavelength director distortions, it may be assumed that the degree of local orientational order is unaffected by the spatial variations of $\mathbf{n}$, so that the equilibrium ODF $\psi_{\rm eq}$ of the cholesteric phase may be assimilated to that $\psi_{\rm eq}^0$ of the uniform nematic state. This approximation has been previously shown to be valid for cholesteric pitches as short as a few dozen particle diameters,\cite{Tort17-1} and is expected to hold without restrictions in our case. $\psi_{\rm eq}$ may then be obtained through the functional minimization of $\mathpzc{f}_0$ (Eq.~\eqref{eq:f0}) in the self-consistent form
\begin{equation*}
  \psi_{\rm eq} (u_x) = \frac{1}{Z} \exp\bigg\{\frac{\rho}{4\pi^2} \int_{-1}^1 du'_x \, \psi_{\rm eq}(u'_x) \kappa(u_x, u'_x) \bigg\},
\end{equation*}
with $Z$ a Lagrange multiplier such that
\begin{equation*}
  4\pi^2 \int_{-1}^1 du_x \, \psi_{\rm eq}(u_x) = 1,
\end{equation*}
and $\kappa$ a generalized excluded-volume kernel,\cite{Fyne98}
\begin{multline}
  \label{eq:kernel}
  \kappa(u_x, u'_x) = \int d\mathbf{r}_{12}  \oiint d \mathcal{R}_1 d\mathcal{R}_2 \, \overline{f}(\mathbf{r}_{12}, \mathcal{R}_1, \mathcal{R}_2)\\
  \times \delta(u_{1x}-u_x)\delta(u_{2x}-u'_x).
\end{multline}
As in Refs.~\onlinecite{Fyne98,vanW12,Tort18}, we sample the conformational distribution $P$ by single-molecule simulations, following the numerical protocol described in the main text (see Materials and Methods). In this context, the Mayer function $\overline{f}$ (Eq.~\eqref{eq:f_ma_ave}) is averaged over all pairs of simulated origami conformations in the computation of Eqs.~\eqref{eq:K2},~\eqref{eq:kt} and~\eqref{eq:kernel}, and the inverse equilibrium cholesteric pitch $q_{\rm eq}$ is finally obtained by minimization of $\mathpzc{f}_d$ at fixed $T$ and $\rho$ (Eq.~\eqref{eq:f_el}),\cite{Stra76} 
\begin{equation*}
  q_{\rm eq}(\rho,T) = \frac{k_t[\psi_{\rm eq}]}{K_2[\psi_{\rm eq}]}.
\end{equation*}

\section{Chiral potential of mean force and phase handedness} \label{app:mean_force}

\begin{figure*}[]
  \includegraphics[width=\textwidth]{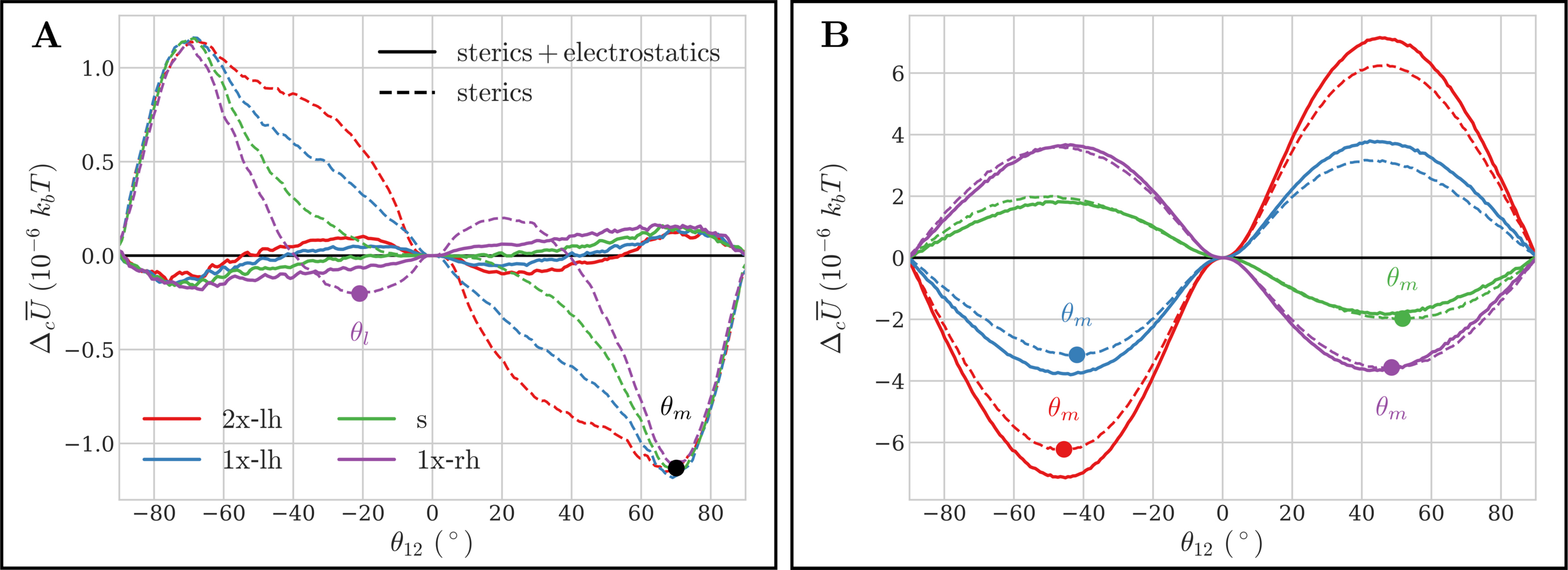}
  \caption{\label{fig1}\textbf{Chiral two-body PMF of ground-state and thermalized origamis.} A) Chiral component of the angular PMF ($\Delta_c \overline{U}$) as a function of origami inter-axis angle ($\theta_{12}$, see Fig.~2A)) for ground-state filaments. Positive (resp.~negative) values of $\theta_{12}$ denote right-handed (resp.~left-handed) two-particle arrangements. Solid dots mark the locations of the curve minima, as discussed in the text, and are only displayed in the case of pure steric repulsion for clarity. B) Same as A) for thermalized origamis.}
\end{figure*}

In the following, let us denote the properties relative to right- and left-handed pair configurations by $+$ and $-$ subscripts, respectively. The angular two-body potential of mean force (PMF) $\overline{U}_\pm$ associated with two-particle arrangements of fixed handedness is given by\cite{Wens11,Wens14}
\begin{equation}
  \label{eq:mean_force}
  \beta \overline{U}_\pm(\theta) \equiv -\log \big\langle e^{-\beta U_{\rm inter}} \big \rangle_\pm^{(\theta)},
\end{equation}
where the configurational average $\big\langle\cdot \big\rangle_\pm^{(\theta)}$ is defined as
\begin{multline}
  \label{eq:mean_ave}
  \!\!\!\!\!\!\big\langle e^{-\beta U_{\rm inter}} \big \rangle_\pm^{(\theta)}  \equiv \frac{1}{V_{\rm int}} \int_V d\mathbf{r}_{12} \oiint d\mathcal{R}_1 d\mathcal{R}_2 \, \delta (\mathbf{u}_1\cdot \mathbf{u}_2 -\cos\theta) \\ \times  \Theta\big\{\pm \mathbf{r}_{12} \cdot (\mathbf{u}_1 \times \mathbf{u}_2)\big\} \, e^{-\beta U_{\rm inter}(\mathbf{r}_{12},  \mathcal{R}_1,  \mathcal{R}_2)} \, ,
\end{multline}
using the notations of Section~\ref{app:FY}. In Eq.~\eqref{eq:mean_ave}, the Heaviside function $\Theta$ mirrors the fact that the handedness of an arrangement of two particles with center-of-mass separation vector $\mathbf{r}_{12} \equiv \mathbf{r}_2-\mathbf{r}_1$ and respective long axes $\mathbf{u}_i \equiv \mathcal{R}_i\cdot\mathbf{e}_x$ is determined by the sign of $ \mathbf{r}_{12} \cdot (\mathbf{u}_1 \times \mathbf{u}_2)$, and $V_{\rm int}$ represents the total volume spanned by the spatial and angular integrals,
\begin{equation*}
  V_{\rm int} = \frac{(8\pi^2)^2}{2}V,
\end{equation*}
where the factor $1/2$ accounts for the equal division of the two-particle configurational space between left- and right-handed arrangements. Note that in the case of flexible particles, Eq.~\eqref{eq:mean_ave} may be further averaged over a representative ensemble of molecular conformations using the numerical procedure outlined in the main text (see Materials and Methods). In this study, we use for the volume $V$ the smallest cubic box containing all possible interacting configurations of any two origami conformations.
\par
In the context of Eqs.~\eqref{eq:mean_force} and \eqref{eq:mean_ave}, a system of two particles with fixed inter-axis angle $\theta_{12}$ (Fig.~2A) may adopt a thermodynamically-stable right-handed configuration if their net repulsion is minimized in a right-handed arrangement --- i.e., if $\overline{U}_+(\theta_{12}) < \overline{U}_-(\theta_{12})$. Conversely, $\overline{U}_+(\theta_{12}) > \overline{U}_-(\theta_{12})$ indicates a thermodynamic preference for left-handed arrangements. The relative stability of chiral two-particle assemblies is thus quantified by the chiral component of the PMF,
\begin{equation}
  \label{eq:chiral_mean}
  \Delta_c \overline{U}(\theta) \equiv \overline{U}_+(\theta) - \overline{U}_-(\theta) = k_b T \log \frac{\big\langle e^{-\beta U_{\rm inter}} \big \rangle_-^{(\theta)}}{\big\langle e^{-\beta U_{\rm inter}} \big \rangle_+^{(\theta)}}.
\end{equation}
In the case of particles with high aspect ratios interacting through short-ranged repulsive potentials, it is easy to verify that only a small statistical fraction of the configurations sampled in Eq.~\eqref{eq:mean_ave} may display a significant interaction energy $U_{\rm inter}>0$, so that
\begin{equation*}
  \big\langle e^{-\beta U_{\rm inter}} \big \rangle_\pm^{(\theta)} \longrightarrow \, 1 \qquad \forall \theta \in [-\pi/2,\pi/2].
\end{equation*}
The Taylor expansion of Eq.~\eqref{eq:chiral_mean} then reads as, to leading order in $1-\big\langle e^{-\beta U_{\rm inter}} \big \rangle_\pm$,
\begin{equation*}
  \Delta_c \overline{U}(\theta) = k_b T \, \Big \{ \big\langle e^{-\beta U_{\rm inter}} \big \rangle_-^{(\theta)} - \big\langle e^{-\beta U_{\rm inter}} \big \rangle_+^{(\theta)} \Big \},
\end{equation*}
and one recovers the definition of the chiral pair excluded volume employed in Refs.~\onlinecite{Duss15} and~\onlinecite{Tort17-2} for systems of hard particles, up to a constant multiplicative prefactor.
\par
It is apparent from Fig.~\ref{fig1} that the chiral PMFs of thermalized origamis are significantly larger in magnitude than those of their respective ground states, and are also relatively insensitive to the inclusion of electrostatic repulsion. These two observations evidence the ascendency of long-wavelength backbone deformations over local axial twist in their LChLC ordering, as the larger lengthscales associated with solenoidal writhe render the chiral assembly of thermalized filaments largely independent of the detailed nature of their much shorter-ranged repulsive interactions. The PMFs of thermalized origamis are further found to bear a unique minimum $\theta_m$ such that $\theta_m < 0$ for left-twisted filaments and $\theta_m > 0$ for their right-twisted counterparts (Fig.~\ref{fig1}B), thus ensuring their stabilization of iso-chiral LChLC arrangements; a thorough discussion of the quantitative link between phase handedness and chiral PMFs may be found in Ref.~\onlinecite{Tort17-2}.
\par
Conversely, the PMFs of ground-state filaments interacting purely through steric repulsion display a shallower minimum at large inter-axis angles ($\theta_m \simeq +\SI{70}{\degree}$, Fig.~\ref{fig1}A), corresponding to the close-approach configuration of ground-state duplexes, as the helical threads of B-DNA form a fixed angle of roughly $\SI{35}{\degree}$ with respect to the normal to the double-helix axis.\cite{Tomb05} This large value is obviously incompatible with the local orientational order of LChLCs, but is nonetheless associated with a regime of weakly-negative values of $\Delta_c \overline{U}$ at smaller angles $\theta_{12}>0$ in the case of the s, 1x-lh and 2x-lh origami variants --- and thus leads to their formation of stable right-handed phases. However, the chiral PMF of 1x-rh filaments bears a local secondary minimum $\theta_l$ at small inter-axis angles of about $-\SI{20}{\degree}$ (Fig.~\ref{fig1}A), arising from their weak right-handed axial twist, which instead stabilizes their left-handed LChLC assembly. 
\par
Finally, we report that electrostatic interactions greatly reduce the magnitude of the chiral PMFs for all ground-state filaments, indicating that the inclusion of longer-ranged repulsion results in an effective screening of their local chiral molecular surfaces --- and therefore unwinds their equilibrium pitches.\cite{Arak01} This conclusion is consistent with the recent results of extensive all-atom simulations of short DNA oligomers,\cite{Cort17} in which the net contribution of electrostatic interactions to the chiral PMF was found to be negligible at comparable monovalent salt concentrations.

\begin{figure*}[]
  \includegraphics[width=\textwidth]{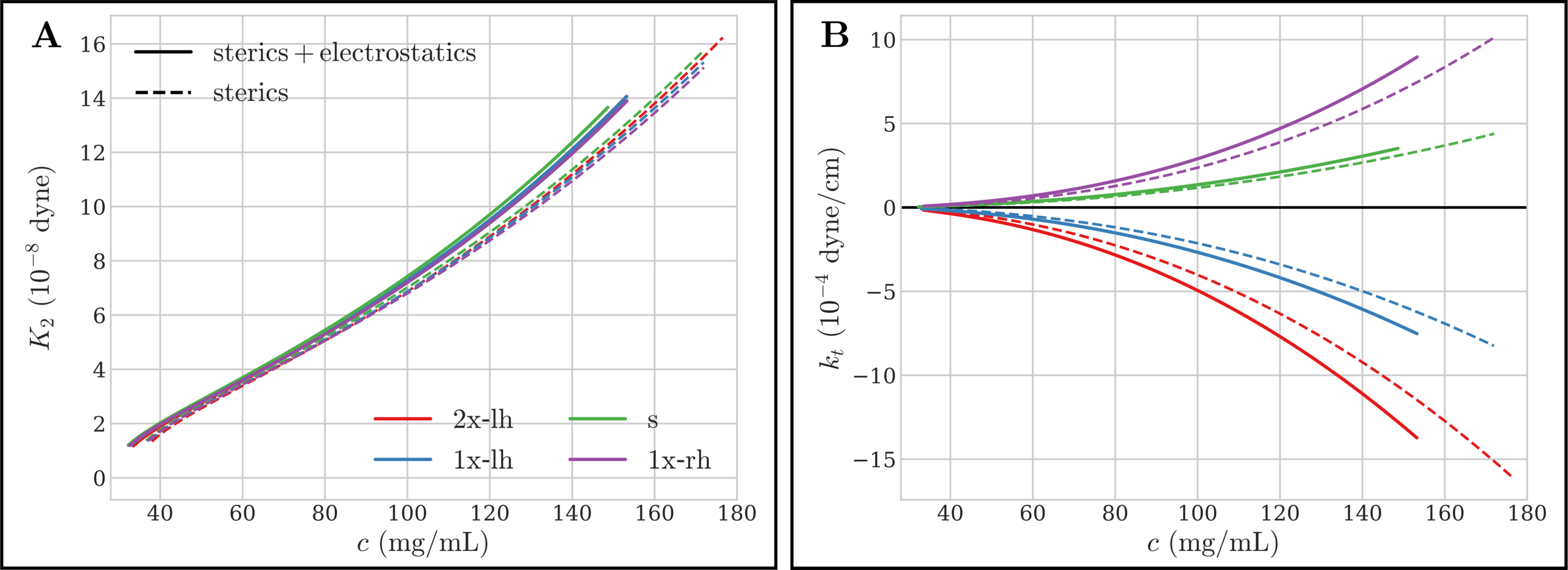}
  \caption{\label{fig2}\textbf{Twist elastic modulus and chiral strength of thermalized origamis.} A) Oseen-Frank twist modulus ($K_2$) as a function of particle concentration ($c$) for the different thermalized origami variants. B) Same as A) for the chiral strength ($k_t$).}
\end{figure*}

\section{Oseen-Frank twist elastic moduli and chiral strengths} \label{app:ops}

We reproduce in Fig.~\ref{fig2} the density dependence of the Oseen-Frank twist elastic modulus $K_2$ and chiral strength $k_t$ in the case of thermalized origami filaments, computed following the procedure outlined in the main text (see Materials and Methods). The orders of magnitudes of the obtained values are in very good agreement with experimental measurements performed in filamentous virus solutions,\cite{Dogi00} whose molecular dimensions, relative flexibility and absolute cholesteric pitches are comparable to those of the origamis.\cite{Siav17} The general tendencies apparent in Fig.~\ref{fig2} are also consistent with experimental results on virus assemblies, with both $K_2$ and $k_t$ displaying a marked increase in magnitude with increasing particle concentration.\cite{Dogi00} The observed stiffening of twist curvature elasticities upon the inclusion of electrostatic repulsion (Fig.~\ref{fig2}A) further mirrors the experimental variations of $K_2$ with decreasing salt concentration in such systems.\cite{Dogi00} The precise experimental determination of these quantities in LChLC phases of origami filaments would be desirable for the thorough investigation of these effects, and for further quantitative comparisons with the theoretical predictions of Fig.~\ref{fig2}.

\section{Derivation of an helicity order parameter} \label{app:helicity}

Let us parametrize an arbitrary backbone conformation of an origami with contour length $l_c$ by a continuous curve $\mathbf{r}(s)$, where $s\in[0,l_c]$ is the curvilinear abscissa. The local unit tangent to the curve reads as
\begin{equation}
  \label{eq:tang}
  \mathbf{t}(s) = \frac{d\mathbf{r}}{ds} \equiv t_\parallel(s) \mathbf{u} + \mathbf{t}_\bot (s),
\end{equation}
with $\mathbf{u}$ the long axis of the conformation as defined in the main text (see Materials and Methods) and $\mathbf{t}_\bot \cdot \mathbf{u} = 0$. Due to the large bending rigidity of the filaments, we assume the transverse fluctuations of $\mathbf{r}$ to be small,
\begin{equation*}
  \lVert \mathbf{t}_\bot(s) \rVert = \bigg \lVert \frac{d\mathbf{r}_\perp}{ds} \bigg \rVert \ll 1,
\end{equation*}
where we used the notation of Fig.~3A in the main text, with $\lVert \cdot \rVert$ the Euclidean norm. Thus,
\begin{equation*}
   t_\parallel(s) = \sqrt{1- \lVert \mathbf{t}_\bot(s) \rVert^2} = 1 + \mathcal{O} \big(\lVert \mathbf{t}_\bot \rVert^2\big),
\end{equation*}
and integrating Eq.~\eqref{eq:tang} yields, to leading order in $\mathbf{t}_\bot$,
\begin{equation}
  \label{eq:r}
  \mathbf{r}(s) = \mathbf{r}_0 + s \mathbf{u} + \mathbf{r}_\bot(s),
\end{equation}
where $\mathbf{r}_0 = \mathbf{r}(0) - \mathbf{r}_\bot(0)$. Consistently with the previous approximations, we further assimilate the filament long axis $\mathbf{u}$ with the normalized end-to-end separation vector,
\begin{equation*}
  \mathbf{u} \cong \frac{\mathbf{r}(l_c)-\mathbf{r}(0)}{\lVert \mathbf{r}(l_c)-\mathbf{r}(0) \rVert},
\end{equation*}
so that Eq.~\eqref{eq:r} imposes simple periodic boundary conditions for $\mathbf{r}_\bot$,
\begin{equation*}
  \mathbf{r}_\bot(0) = \mathbf{r}_\bot(l_c).
\end{equation*}
\par
$\mathbf{r}_\bot$ may then be expressed in the form of an inverse Fourier transform,
\begin{equation}
  \label{eq:inv_trans}
  \mathbf{r}_\bot(s) = \frac{1}{l_c} \sum_k \, \widehat{\mathbf{r}}_\bot (k) \times e^{2i\pi ks},
\end{equation}
with discrete wavenumbers $k=n/l_c$ for any non-zero integer $n$ and coefficients
\begin{align}
  \label{eq:fourier_proj}
  \widehat{\mathbf{r}}_\bot (k) &=  \int_0^{l_c} ds \, \mathbf{r}_\bot (s) \times e^{-2i\pi ks} \nonumber \\
                                                &\equiv \widehat{r}_{\bot v}(k) \mathbf{v} +\widehat{r}_{\bot w}(k) \mathbf{w}.
\end{align}
Let $\abs{\cdot}$ be the complex modulus, and
\begin{align}
  \label{eq:phi_v}
  e^{i \phi_v(k)} &\equiv \frac{\widehat{r}_{\bot v}(k)}{\abs{\widehat{r}_{\bot v}(k)}}, \\
  \label{eq:phi_w}
  e^{i\phi_w(k)} &\equiv \frac{\widehat{r}_{\bot w}(k)}{\abs{\widehat{r}_{\bot w}(k)}}.
 \end{align} 
Using Eqs.~\eqref{eq:r}--\eqref{eq:fourier_proj}, the backbone conformation $\mathbf{r}_k$ associated with a transverse deformation mode of arbitrary wavenumber $k$ is given by the parametric equation
\begin{multline}
  \label{eq:ell_helix}
  \mathbf{r}_k(s) = s\mathbf{u} + \frac{2}{l_c} \Big \{ \abs{\widehat{r}_{\bot v}(k)} \cos[2\pi ks + \phi_v(k)] \mathbf v \\+ \abs{\widehat{r}_{\bot w}(k)}\cos[2\pi ks + \phi_w(k)] \mathbf w \Big \}.
\end{multline}
\par
In the most general case, Eq.~\eqref{eq:ell_helix} describes an elliptical helix of axis $\mathbf{u}$ and pitch $p=1/k$. The shape chirality associated with a deformation mode $\mathbf{r}_k$ is thus quantified by the anisotropy of its elliptical cross-section, which we now proceed to analyse. In the following, we omit some of the explicit $k$ dependences in order to alleviate the notations when no confusion can arise. Let us denote by
\begin{equation}
  \label{eq:hermit}
   \lVert \widehat{\mathbf{r}}_\bot \rVert \equiv \sqrt{\widehat{\mathbf{r}}_\bot \cdot \widehat{\mathbf{r}}^{\:*}_\bot} = \sqrt{\abs{\widehat{r}_{\bot v}}^2+\abs{\widehat{r}_{\bot w}}^2}
\end{equation}
the Euclidean modulus of $\widehat{\mathbf{r}}_\bot$, and define
\begin{align}
  \label{eq:theta}
  \theta& \equiv \arccos\frac{\abs{\widehat{r}_{\bot v}}}{\lVert \widehat{\mathbf{r}}_\bot \rVert}, \\
  \label{eq:A}
  A &\equiv \frac{2\lVert \widehat{\mathbf{r}}_\bot \rVert}{l_c}.
\end{align}
Using Eqs.~\eqref{eq:ell_helix}--\eqref{eq:A}, the transverse components of $\mathbf{r}_k$ may be rewritten as
\begin{gather}
  \label{eq:rv}
  r_{kv}(s)   \equiv \mathbf{r}_k(s)\cdot \mathbf{v} =  A \cos\theta \times \cos(\omega s+\phi_v), \\
  \label{eq:rw}
  r_{kw}(s)  \equiv \mathbf{r}_k(s)\cdot \mathbf{w}=  A \sin\theta \times \cos(\omega s+\phi_w),
\end{gather}
with $\omega \equiv 2\pi k$. Eq.~\eqref{eq:rw} then yields
\begin{equation*}
  \frac{r_{kw}(s)}{A \sin\theta} = \cos(\omega s + \phi_v) \cos \phi +\sin(\omega s + \phi_v) \sin \phi,
\end{equation*}
where 
\begin{equation}
  \label{eq:phi}
  \phi\equiv \phi_v-\phi_w. 
\end{equation}
Thus, using Eq.~\eqref{eq:rv},
\begin{equation}
  \label{eq:implicit_1}
  \frac{r_{kw}(s)}{A \sin\theta}-\frac{r_{kv}(s)}{A \cos\theta} \cos \phi = \sin(\omega s + \phi_v) \sin \phi,
\end{equation}
and Eq.~\eqref{eq:rv} immediately yields the further relation
\begin{equation}
  \label{eq:implicit_2}
  \sin(\omega s + \phi_v) = \pm \sqrt{1-\bigg\{\frac{r_{kv}(s)}{A\cos\theta}\bigg\}^2}.
\end{equation}
Plugging Eq.~\eqref{eq:implicit_2} into Eq.~\eqref{eq:implicit_1} leads to a quadratic equation for $r_{kv}$ and $r_{kw}$,
\begin{equation}
  \label{eq:implicit_3}
  \bigg(\frac{r_{kv}}{A\cos\theta}\bigg)^2+\bigg(\frac{r_{kw}}{A\sin\theta}\bigg)^2 - 2\cos\phi \frac{r_{kv} r_{kw}}{A^2\cos\theta\sin\theta} = \sin^2\phi.
\end{equation}
Denoting by $\mathbf{r}_{k\bot}$ the total transverse component of $\mathbf{r}_k$,
\begin{equation*}
 \mathbf{r}_{k\bot}(s) \equiv r_{kv}(s)\mathbf{v}+ r_{kw}(s)\mathbf{w}, 
 \end{equation*}
Eq.~\eqref{eq:implicit_3} may be recast in the compact form
\begin{equation*}
  \mathbf{r}_{k\bot}^{\sf T} \cdot \mathcal{Q} \cdot \mathbf{r}_{k\bot} = 1,
\end{equation*}
with $\mathbf{r}_{k\bot}^{\sf T}$ the matrix transpose of $\mathbf{r}_{k\bot}$ and $\mathcal{Q}$ the matrix representation of the quadratic form in Eq.~\eqref{eq:implicit_3},
\begin{align*}
\mathcal{Q} &= \frac{1}{\sin^2 \phi}
\begin{bmatrix}
\frac{\displaystyle1}{\displaystyle(A\cos\theta)^2} & -\frac{\displaystyle \cos\phi}{\displaystyle A^2\cos\theta\sin\theta} \\
 -\frac{\displaystyle\cos\phi}{\displaystyle A^2\cos\theta\sin\theta} & \frac{\displaystyle1}{\displaystyle(A\sin\theta)^2}
\end{bmatrix}.
\end{align*}
The respective lengths $r_\pm$ of the semi-major and semi-minor elliptical axes are then related to the respective largest and smallest eigenvalues $\lambda_\pm$ of $\mathcal{Q}$ through\cite{Salm00}
\begin{equation*}
r_\pm = 1/\sqrt{\lambda_\mp},
\end{equation*}
which yields, after rearrangements,
\begin{equation}
  \label{eq:eccentricity}
  r_\pm = A \sqrt{\frac{1\pm \sqrt{1-\sin^2\phi\sin^2 2\theta}}{2}}.
\end{equation}
Interestingly, Eq.~\eqref{eq:eccentricity} bears a strong resemblance to the Jones vector parametrization of the polarization ellipse in classical electrodynamics,\cite{Jone41} which stems from the similarity between Eq.~\eqref{eq:ell_helix} and the field equation of a polarized electromagnetic wave propagating along the direction $\mathbf{u}$. 
\par
Let us define 
\begin{equation}
  \label{eq:h_1}
  \mathcal{H} \equiv \sin \phi\sin 2\theta.
\end{equation}
An explicit expression for $\mathcal{H}$ in terms of the Fourier components $\widehat{\mathbf{r}}_\bot (k)$ may be obtained by substituting Eqs.~\eqref{eq:hermit} and \eqref{eq:theta} for $\theta$,
\begin{equation*}
  \sin2\theta = 2\cos\theta\sin\theta = \frac{2\abs{\widehat{r}_{\bot v}} \abs{\widehat{r}_{\bot w}}}{\abs{\widehat{r}_{\bot v}}^2+ \abs{\widehat{r}_{\bot w}}^2},
\end{equation*}
and substituting Eqs.~\eqref{eq:phi},~\eqref{eq:phi_v} and \eqref{eq:phi_w} for $\phi$,
\begin{equation*}
  e^{i\phi} = e^{i\phi_v} e^{-i\phi_w} = \frac{\widehat{r}_{\bot v} \times \widehat{r}^{\:*}_{\bot w}}{\abs{\widehat{r}_{\bot v}} \abs{\widehat{r}_{\bot w}}}.
\end{equation*}
Eq.~\eqref{eq:h_1} may thus be rewritten in the form
\begin{equation}
  \label{eq:h_2}
  \mathcal{H} = 2\times \frac{\Im \big\{\widehat{r}_{\bot v} \times \widehat{r}^{\:*}_{\bot w}\big\}}{\abs{\widehat{r}_{\bot v}}^2+ \abs{\widehat{r}_{\bot w}}^2} =  2 \times\frac{\Im \big \{ \widehat{c}_{vw} \big \} }{\widehat{c}_{vv} + \widehat{c}_{ww}},
\end{equation}
and one recovers the definition of Eq.~4 in the main text, with $\widehat{c}_{vw}(k)$ the Fourier components of the cross-correlation function of $r_{\bot v}$ and $r_{\bot w}$ as given by the convolution theorem,
\begin{equation*}
  \widehat{c}_{vw}(k) = \widehat{r}_{\bot v}(k) \times \widehat{r}^{\:*}_{\bot w}(k).
\end{equation*}
\par
Using Eqs.~\eqref{eq:eccentricity} and \eqref{eq:h_1}, the transverse eccentricity of the elliptical cross-section reads as
\begin{equation*}
  e^2 \equiv \frac{r_+^2 - r_-^2}{r_+^2} = \frac{2\sqrt{1-\mathcal{H}^2}}{1+\sqrt{1-\mathcal{H}^2}}.
\end{equation*}
A necessary and sufficient condition for the deformation mode $\mathbf{r}_k$ to describe an ideal circular helix is given by
\begin{equation*}
  e(k) = 0 \iff \mathcal{H}(k) = \pm 1.
\end{equation*}
Conversely, 
\begin{equation*}
  e(k) = 1 \iff \mathcal{H}(k) = 0
\end{equation*}
describes the degenerate case in which the elliptical cross-section collapses to a flat line segment, leading to an achiral deformation mode. The mean solenoidal radius $r_m$ of an arbitrary deformation mode may finally be obtained in the compact form
\begin{equation*}
  r_m \equiv \sqrt{r_+ \times r_-} = A \sqrt{\frac{\abs{\mathcal{H}}}{2}}.
\end{equation*}
\par 
The magnitude of $\mathcal{H}(k)$ may thus be understood as a measure of the degree of circular helicity of the deformation mode $\mathbf{r}_k$. The link between the sign of $\mathcal{H}(k)$ and the corresponding helical handedness may be elucidated by considering the case of an ideal circular helical conformation of axis $\mathbf{u}$, radius $r_h>0$ and inverse pitch $q=1/p_h$. The general parametric equation of such a conformation reads as, in the limit of weak helical curvature ($q r_h \ll 1$),
\begin{equation}
  \label{eq:hel}
  \mathbf{r}^h_q(s) = s\mathbf{u} + r_h\cos(2\pi q s + \phi_h) \mathbf{v} + r_h \sin (2\pi q s + \phi_h) \mathbf{w},
\end{equation}
with $\phi_h \in [0,2\pi]$. In the convention of Eq.~\eqref{eq:hel}, the handedness of the helix is quantified by the sign of $q$, with $q>0$ (resp.~$q<0$) corresponding to a right-handed (resp.~left-handed) helicity. Using the previous notations, the Fourier components of the transverse vector $\mathbf{r}^h_{q\bot}$ associated with Eq.~\eqref{eq:hel} read as
\begin{align*}
  \widehat{\mathbf{r}}^{\:h}_{q\bot}(k) = 
    \begin{dcases}
     l_c \times \frac{r_h e^{\pm i\phi_h}}{2} \big(\mathbf{v} \pm e^{-i\pi/2} \mathbf{w} \big )  &\text{if } \hphantom{|\,} k\hphantom{|} = \pm q \\
     \mathbf{0}  &\text{if } \abs{k} \neq \abs{q}
   \end{dcases}.
\end{align*}
In this case, for any wavenumber $k>0$, Eq.~\eqref{eq:h_2} reduces to
\begin{equation*}
  \mathcal{H}(k) =  \delta_{k,\abs{q}} \times \sgn q,
\end{equation*}
and it is easy to check that Eqs.~\eqref{eq:eccentricity},~\eqref{eq:hermit} and \eqref{eq:A} yield
\begin{equation*}
  r_+(k)=r_-(k)= \delta_{k,\abs{q}} \times r_h,
\end{equation*} 
with $\delta$ the Kronecker delta and $\sgn$ the sign function. Therefore, the handedness of a deformation mode with arbitrary wavenumber $k>0$ may be determined by the sign of $\mathcal{H}(k)$, with $\mathcal{H}(k)> 0$ and $\mathcal{H}(k)< 0$ respectively describing a right- and left-handed helicity.

\section{Fluctuation spectrum from the equipartition theorem} \label{app:equipartition}

Using the notations of Section~\ref{app:helicity}, the enthalpic penalty associated with the bending response of a single origami to thermal fluctuations reads as, in the case of weak curvature deformations,
\begin{equation}
  \label{eq:hamiltonian_1}
  \Delta H_{\rm bend} = \frac{\mathpzc{K}}{2} \int_0^{l_c} ds \, \bigg\lVert  \frac{d^2\mathbf{r}_\bot}{ds^2} \bigg \rVert^2,
\end{equation}
where the bending modulus $\mathpzc{K}$ is related to the origami persistence length $l_p$ through
\begin{equation}
  \label{eq:kappa}
  \mathpzc{K} = l_p k_b T.
\end{equation}
Substituting Eqs.~\eqref{eq:inv_trans} and \eqref{eq:fourier_proj} for $\mathbf{r}_\bot$ in Eq.~\eqref{eq:hamiltonian_1} yields
\begin{equation}
  \label{eq:hamiltonian_2}
  \Delta H_{\rm bend} = \frac{\mathpzc{K}}{2l_c} \times \sum_k (2\pi k)^4 \Big \{ \abs{\widehat{r}_{\bot v}(k)}^2 + \abs{\widehat{r}_{\bot w}(k)}^2 \Big \}.
\end{equation}
Assimilating the different transverse deformation modes in Eq.~\eqref{eq:hamiltonian_2} to decoupled degrees of freedom, the equipartition theorem imposes for $\widehat{r}_{\bot v}$ and $\widehat{r}_{\bot w}$
\begin{equation*}
  \big \langle\abs{\widehat{r}_{\bot v}(k)}^2 \big \rangle =  \big \langle\abs{\widehat{r}_{\bot w}(k)}^2\big \rangle = \frac{k_b T l_c}{\mathpzc{K}} \times \frac{1}{(2\pi k)^4}.
\end{equation*}
Thus, using Eqs.~\eqref{eq:hermit} and \eqref{eq:kappa},
\begin{equation}
   \big \langle \lVert \widehat{\mathbf{r}}_\bot(k) \rVert^2 \big \rangle = \frac{l_c}{l_p} \times \frac{1}{8 \pi^4 k^4},
\end{equation}
 valid in the limit of long-wavelength fluctuations ($k \to 0$).

 \section{Twist-writhe conversion and helical fluctuations} \label{app:twist_writhe}
Let us consider a long origami filament whose extremities are firmly clamped to impose the parallel alignment of its backbone end tangents,
\begin{equation}
  \label{eq:clamped}
   \frac{d\mathbf{r}}{ds}\bigg|_{s=0} = \frac{d\mathbf{r}}{ds}\bigg|_{s=l_c} \equiv \mathbf{t}_0.
\end{equation}
The origami backbone curve $\mathbf{r}$ is defined as
\begin{equation*}
  \mathbf{r} \equiv \frac{1}{6} \sum_{i=1}^6 \mathbf{r}_i,
\end{equation*}
where the continuous centerline $\mathbf{r}_i(s_i)$ of the $i$-th constituent DNA duplex is obtained by contour interpolation of the center-of-mass positions of its bonded nucleotides (see Materials and Methods). For simplicity, we neglect the effects of duplex splaying at the origami ends, and thus assume Eq.~\eqref{eq:clamped} to hold at each of the center curve extremities,
\begin{equation*}
  \frac{d\mathbf{r}_i}{ds_i}\bigg|_{s_i=0} = \frac{d\mathbf{r}_i}{ds_i}\bigg|_{s_i=l_i} = \mathbf{t}_0.
\end{equation*}
We further restrict our study to the regime of weak bending deformations of the duplex centerlines about the straight backbone conformation of the origami ground state, and neglect potential fluctuations in their respective contour lengths $l_i$.
\par
Under these assumptions, the formulation of the C\u{a}lug\u{a}reanu-Fuller-White theorem extended to the treatment of open curves\cite{Berg06} states that the linking number ${\rm Lk}_i$ of each individual duplex may be decomposed into twist and writhe contributions,
\begin{equation}
  \label{eq:cfw}
  {\rm Lk}_i = {\rm Tw}_i + {\rm Wr}_i.
\end{equation}
In this context, ${\rm Lk}_i$ represents the (signed) number of net right-handed turns per unit contour length by which the two strands of the duplex wind around $\mathbf{t}_0$.\cite{Berg06} These turns may result in both a local twist of the strands about their common centerline $\mathbf{r}_i$, as quantified by the twist density ${\rm Tw}_i$, and/or in a global supercoiling of the centerline itself, as measured by the writhe integral ${\rm Wr}_i$. It should be noted that the linking number ${\rm Lk}_i$ is generally not a topological invariant in the case of non-circular DNA fragments. Within the origami filament architecture, ${\rm Lk}_i$ is initially constrained by the designed locations of the inter-helical crossovers, but may partially relax towards its preferred unhindered value ${\rm Lk}_0$ --- thus inducing global axial twist in the origami ground state.\cite{Diet09}
\par
Within ground-state B-DNA, the relaxed linking number ${\rm Lk}_0$ is entirely absorbed in the form of twist strain,
\begin{equation}
  \label{eq:lk0}
  {\rm Lk}_0 = {\rm Tw}_0 \simeq \SI{1/10.5}{\per\bp}.
\end{equation}
The axial twist handedness of an origami filament comprised of duplexes with linking number ${\rm Lk}_i$ is therefore determined by the sign of $\Delta{\rm Lk}_i \equiv {\rm Lk}_i - {\rm Lk}_0$, with $\Delta{\rm Lk}_i>0$ ($\Delta{\rm Lk}_i<0$) respectively denoting a residual over-winding (under-winding) of the duplexes, associated with a global left-handed (right-handed) compensatory twist of the origami. The total elastic energy of a constituent duplex, as defined by an arbitrary centerline curve $\mathbf{r}_i$ and uniform twist density ${\rm Tw}_i$, may be obtained as a straightforward generalization of Eq.~\eqref{eq:hamiltonian_1},\cite{Mark95}
\begin{equation*}
  \Delta H_i = \frac{1}{2} \int_0^{l_i} ds_i \, \bigg\{ \mathpzc{K}_i \bigg\lVert  \frac{d^2\mathbf{r}_i}{ds_i^2} \bigg \rVert^2 + 4\pi^2C_i \Big({\rm Tw}_i-{\rm Tw}_0\Big)^2\bigg\},
\end{equation*}
with $\mathpzc{K}_i$ and $C_i$ the respective effective bending and twisting moduli of B-DNA within the origami structure. Eqs.~\eqref{eq:cfw} and \eqref{eq:lk0} immediately yield
\begin{equation}
  \label{eq:h_i}
  \Delta H_i =  \frac{\mathpzc{K}_i}{2} \int_0^{l_i} ds_i \, \bigg\lVert  \frac{d^2\mathbf{r}_i}{ds_i^2} \bigg \rVert^2 + 2\pi^2 C_i l_i \Big(\Delta{\rm Lk}_i-{\rm Wr}_i\Big)^2.
\end{equation}
\par
It is apparent that the twist elastic contribution in Eq.~\eqref{eq:h_i} is minimized by conformations in which $\Delta{\rm Lk}_i$ and ${\rm Wr}_i$ bear equal sign and magnitude, leading to a favored positive (right-handed) supercoiling in the case of left-twisted origamis ($\Delta{\rm Lk}_i>0$, ${\rm Wr}_i>0$), and negative (left-handed) supercoiling for their right-handed counterparts ($\Delta{\rm Lk}_i<0$, ${\rm Wr}_i<0$). However, this twist relaxation mechanism is hindered by the high penalty in bending energy arising from the finite curvature of the resulting solenoidal centerline deformations. The competition of these two effects, acting constructively on each duplex within the origami structures, leads to the weak anti-chiral backbone fluctuations underpinning their LChLC assembly.


\bibliography{refs}
\bibliographystyle{Science}